\documentclass[twocolumn]{aastex62}

\usepackage{longtable}

\received{\today}
\revised{}
\accepted{}

\submitjournal{ApJ}

\shorttitle{Energetic and asymmetric SN~2017gmr }
\shortauthors{Andrews et al.}

\begin{document}

\title{SN~2017gmr: An energetic Type II-P supernova with asymmetries}

\correspondingauthor{Jennifer Andrews}
\email{jandrews@as.arizona.edu}

\newcommand{\UA}{\affiliation{Steward Observatory, University of Arizona, 933 North Cherry Avenue, Tucson, AZ 85721-0065, USA}}

\newcommand{\UCDavis}{\affiliation{Department of Physics, University of California, 1 Shields Avenue, Davis, CA 95616-5270, USA}}

\newcommand{\ARI}{\affiliation{Aryabhatta Research Institute of observational sciences, Manora Peak, Nainital 263 001, India}}

\newcommand{\IIA}{\affil{Indian Institute of Astrophysics, Koramangala, Bengaluru 560 034, India}}

\newcommand{\UCSB}{\affil{Department of Physics, University of California, Santa Barbara, CA 93106-9530, USA}}

\newcommand{\LCO}{\affil{Las Cumbres Observatory, 6740 Cortona Dr, Suite 102, Goleta, CA 93117-5575, USA}}

\newcommand{\TAM}{\affil{Mitchell Institute for Fundamental Physics and Astronomy, Texas A\&M University,College Station, TX 77843, USA}}

\newcommand{\CfA}{\affil{Center for Astrophysics \textbar{} Harvard \& Smithsonian, 60 Garden Street, Cambridge, MA 02138-1516, USA}}

\newcommand{\aarhus}{\affil {Department of Physics and Astronomy, Aarhus University, Ny Munkegade 120, DK-8000 Aarhus C, Denmark}}

\newcommand{\JAP}{\affil {Joint Astronomy Programme, Department of Physics, Indian Institute of Science, Bengaluru 560012, India}}

\newcommand{\Dublin}{\affil {School of Physics, O'Brien Centre for Science North, University College Dublin, Belfield, Dublin 4, Ireland}}

\newcommand{\Turku}{\affil {Tuorla Observatory, Department of Physics and Astronomy, University of Turku, FI-20014 Turku, Finland}}

\newcommand{\Aryabhatta}{\affil{Aryabhatta Research Institute of Observational Sciences, Manora Peak, Nainital 263 001, India}}

\newcommand{\RaviShankar}{\affil {Pt. Ravi Shankar Shukla University, Raipur 492 010, India}}

\newcommand{\Atacama}{\affil {Instituto de Astronom\'{\i}a y Ciencias Planetarias, Universidad de Atacama, Copayapu 485, Copiap\'o, Chile}}

\newcommand{\MIAChile}{\affil {Millennium Institute of Astrophysics (MAS), Nuncio Monse\~nor S\'otero Sanz 100, Providencia, Santiago, Chile}}

\newcommand{\Warsaw}{\affil {Warsaw University Astronomical Observatory, Al. Ujazdowskie 4, 00-478 Warszawa, Poland}}

\newcommand{\OskarKlein}{\affil {The Oskar Klein Centre, Physics Department, Stockholm University, SE-106 91 Stockholm, Sweden}}

\newcommand{\MSU} {\affil{Department of Physics and Astronomy, Michigan State University,East Lansing, MI 48824, USA}}

\newcommand{\Belfast}{\affil{Astrophysics Research Centre, School of Mathematics and Physics,Queen’s University Belfast, BT7 1NN, UK}}

\newcommand{\Bello}{\affil{Departamento de Ciencias Fisicas, Universidad Andres Bello, Avda. Republica 252, Santiago, Chile}}

\newcommand{\LasCampanas}{\affil {Carnegie Observatories, Las Campanas Observatory, Casilla 601, La Serena, Chile}}

\newcommand{\Trinity}{\affil {School of Physics, Trinity College Dublin, Dublin 2, Ireland}}

\newcommand{\Edinburgh}{\affil {Institute for Astronomy, University of Edinburgh, Royal Observatory, Blackford Hill, Edinburgh EH9 3HJ, UK}}

\newcommand{\TelAviv}{\affiliation{The School of Physics and Astronomy, Tel Aviv University, Tel Aviv 69978, Israel}}

\newcommand{\INAF}{\affil {INAF - Osservatorio Astronomico di Padova, Vicolo dell'Osservatorio 5, I-35122 Padova, Italy}}

\newcommand{\WeizInst}{\affil {Department of Particle Phys., Astrophys., Weizmann Institute of Science, Rehovot 76100, Israel}}

\newcommand{\PITT}{\affil {PITT PACC, Department of Physics and Astronomy, University of Pittsburgh, Pittsburgh, PA 15260, USA}}

\newcommand {\Stockholm} {\affil{Department of Astronomy and The Oskar Klein Centre, AlbaNova University Centre, Stockholm University, SE-106 91 Stockholm, Sweden}}

\newcommand{\FSU}{ \affiliation{Department of Physics, Florida State University, Tallahassee, FL 32306, USA}}

\newcommand{\UT}{\affiliation{Department of Astronomy, University of Texas at Austin, Austin, TX 78712, USA}}

\newcommand{\Cardiff}{\affiliation{School of Physics and Astronomy, Cardiff University, Queens Buildings, The Parade, Cardiff, CF24 3AA, UK}}

\newcommand{\Konkoly}{\affiliation{Konkoly Observatory, Research Centre for Astronomy and Earth Sciences, Hungarian Academy of Sciences, H-1121 Budapest, Hungary}}

\newcommand{\ESO}{\affil{European Southern Observatory, Karl-Schwarzschild-Str. 2, 85748 Garching b. M\"{u}nchen, Germany}}

\newcommand{\Beijing}{\affil{Key Laboratory of Optical Astronomy, National Astronomical Observatories, Chinese Academy of Sciences, Beijing 100101, China}}

\newcommand{\ChinaChile}{\affil{Chinese Academy of Sciences South America Center for Astronomy, China-Chile Joint Center for Astronomy, Camino El Observatorio \#1515, Las Condes, Santiago, Chile}}

\newcommand{\UNC}{\affiliation{Department of Physics and Astronomy, University of North Carolina at Chapel Hill, Chapel Hill, NC 27599, USA}}

\newcommand{\ICE}{\affil {Institute of Space Sciences (ICE, CSIC), Campus UAB, Carrer de Can Magrans s/n, 08193 Barcelona, Spain
}}

\newcommand{\IEEC}{\affil {Institut d’Estudis Espacials de Catalunya (IEEC), c/Gran Capit\'a 2-4, Edif. Nexus 201, 08034 Barcelona, Spain
}}

\newcommand{\Tsinghua}{\affil{Physics Department and Tsinghua Center for Astrophysics (THCA), Tsinghua University, Beijing, 100084, China}}

\newcommand{\NOAC}{\affil{Key Laboratory of Optical Astronomy, National Astronomical Observatories, Chinese Academy of Sciences, Beijing 100101,  China}}

\newcommand{\Pulkova}{\affiliation{Central (Pulkovo) Observatory of Russian Academy of Sciences, 196140 Pulkovskoye Ave. 65/1, Saint Petersburg, Russia}}

\newcommand{\Racah}{\affil{Racah Institute of Physics, The Hebrew University of Jerusalem, Jerusalem 91904, Israel}}

\newcommand{\Liverpool}{\affil{Astrophysics Research Institute, Liverpool John Moores University, IC2, Liverpool Science Park, 146 Brownlow Hill, Liverpool L3 5RF, UK}}

\newcommand{\Tokyo}{\affil{Division of Theoretical Astronomy, National Astronomical Observatory of Japan, National Institutes of Natural Sciences, 2-21-1 Osawa, Mitaka, Tokyo 181-8588, Japan}}

\newcommand{\LBL}{\affil{Physics Division, Lawrence Berkeley National Laboratory, 1 Cyclotron Road, Berkeley, CA 94720, USA}}

\newcommand{\Aalto}{\affil{Aalto University Mets{\"a}hovi Radio Observatory, Mets{\"a}hovintie 114, FI-02540 Kylm{\"a}l{\"a}, Finland}}

\newcommand{\Chinacfa}{\affil{Center for Astrophysics, Guangzhou University, Guangzhou 510006, China}}

\newcommand{\MPI}{\affil{Max-Planck-Institut fur Astrophysik, Karl-Schwarzschild-Str. 1, D-85748 Garching, Germany}}

\newcommand{\DARK}{\affil{DARK, Niels Bohr Institute, University of Copenhagen, Lyngbyvej 2, 2100 Copenhagen, Denmark}}

\newcommand{\FINCA}{\affil { Finnish Centre for Astronomy with ESO (FINCA), University of Turku, Quantum, Vesilinnantie 5, 20014 University of Turku, Finland}}

\newcommand{\Delhi}{\affil {Department of Physics $\&$ Astrophysics, University of Delhi, Delhi-110 007, India}}
\author[0000-0003-0123-0062]{Jennifer~E. Andrews}
\UA

\author[0000-0003-4102-380X]{D.~J. Sand}
\UA

\author[0000-0001-8818-0795]{S. Valenti}
\UCDavis

\author{Nathan Smith}
\UA

\author[0000-0001-6191-7160]{Raya Dastidar}
\Aryabhatta
\Delhi

\author{D.K. Sahu}
\IIA

\author[0000-0003-1637-267X]{Kuntal Misra}
\ARI
\UCDavis

\author[0000-0003-2091-622X]{Avinash Singh}
\IIA
\JAP

\author[0000-0002-1125-9187]{D. Hiramatsu}
\UCSB
\LCO

\author[0000-0001-6272-5507]{P.J. Brown}
\TAM

\author[0000-0002-0832-2974]{G. Hosseinzadeh}
\CfA

\author{S. Wyatt}
\UA

\author[0000-0001-8764-7832]{J. Vinko}
\Konkoly
\UT

\author{G.C. Anupama}
\IIA

\author[0000-0001-7090-4898]{I. Arcavi}
\TelAviv

\author[0000-0002-5221-7557]{Chris Ashall}
\FSU

\author{S. Benetti}
\INAF

\author{Marco Berton}
\Turku
\Aalto

\author[0000-0002-4294-444X]{K.~A. Bostroem}
\UCDavis 

\author{M. Bulla}
\OskarKlein

\author{J. Burke}
\UCSB
\LCO

\author{S. Chen}
\INAF
\Chinacfa

\author{L. Chomiuk}
\MSU

\author{A. Cikota}
\LBL

\author{E. Congiu}
\INAF
\LCO

\author{B. Cseh}
\Konkoly

\author{Scott Davis}
\FSU

\author{N. Elias-Rosa}
\ICE
\IEEC

\author{T. Faran}
\Racah

\author[0000-0003-2191-1674]{Morgan Fraser}
\Dublin

\author{L. Galbany}
\PITT

\author{C. Gall}
\DARK

\author{A. Gal-Yam}
\WeizInst

\author{Anjasha Gangopadhyay}
\Aryabhatta
\RaviShankar

\author{M. Gromadzki}
\Warsaw

\author{J. Haislip}
\UNC

\author[0000-0003-4253-656X]{D. A. Howell}
\UCSB
\LCO

\author[0000-0003-1039-2928]{E.~Y.  Hsiao}
\FSU

\author{C. Inserra}
\Cardiff

\author[0000-0001-8257-3512]{E. Kankare}
\Turku

\author{H. Kuncarayakti}
\FINCA
\Turku

\author{V. Kouprianov}
\UNC
\Pulkova

\author[0000-0001-7225-2475]{Brajesh Kumar}
\IIA

\author{Xue Li}
\Tsinghua

\author{Han Lin}
\Tsinghua

\author[0000-0002-9770-3508]{K. Maguire}
\Trinity

\author{P. Mazzali}
\Liverpool
\MPI

\author[0000-0001-5807-7893]{C. McCully}
\UCSB
\LCO

\author{P. Milne}
\UA

\author{Jun Mo}
\Tsinghua

\author[0000-0003-2535-3091]{N. Morrell}
\LasCampanas

\author[0000-0002-2555-3192]{M. Nicholl}
\Edinburgh

\author{P. Ochner}
\INAF

\author{F. Olivares}
\Atacama
\MIAChile

\author{A. Pastorello}
\INAF

\author{F. Patat}
\ESO

\author[0000-0003-2734-0796]{M. Phillips}
\LasCampanas

\author[0000-0003-4254-2724]{G. Pignata}
\MIAChile
\Bello

\author{S. Prentice}
\Trinity

\author[0000-0003-4254-2724]{A. Reguitti}
\MIAChile
\Bello

\author[0000-0002-5060-3673]{D. E. Reichart}
\UNC

\author[0000-0001-8651-8772]{\'O. Rodr\'iguez}
\MIAChile
\Bello

\author{Liming Rui}
\Tsinghua

\author{Pankaj Sanwal}
\ARI
\RaviShankar

\author{K. S\'arneczky}
\Konkoly

\author[0000-0002-9301-5302]{M. Shahbandeh}
\FSU

\author{Mridweeka Singh}
\Aryabhatta
\RaviShankar

\author{S. Smartt}
\Belfast

\author{J. Strader}
\MSU

\author[0000-0002-5571-1833]{M.D. Stritzinger}
\aarhus

\author{R. Szak\'ats}
\Konkoly

\author{L. Tartaglia}
\Stockholm

\author{Huijuan Wang}
\Beijing

\author{Lingzhi Wang}
\Beijing
\ChinaChile

\author[0000-0002-7334-2357]{Xiaofeng Wang}
\Tsinghua

\author[0000-0003-1349-6538]{J. C. Wheeler}
\UT
\author{Danfeng Xiang}
\Tsinghua

\author{O. Yaron}
\WeizInst

\author[0000-0002-1229-2499]{D.R. Young}
\Belfast

\author{Junbo Zhang}
\Beijing

\begin{abstract}
We present high-cadence ultraviolet (UV), optical, and near-infrared (NIR) data on the luminous Type II-P supernova SN~2017gmr from hours after discovery through the first 180 days.  SN~2017gmr does not show signs of narrow, high-ionization emission lines in the early optical spectra, yet the optical lightcurve evolution suggests that an extra energy source from circumstellar medium (CSM) interaction must be present for at least 2 days after explosion. Modeling of the early lightcurve indicates a $\sim$ 500R$_{\sun}$ progenitor radius, consistent with a rather compact red supergiant, and late-time luminosities indicate up to 0.130 $\pm$ 0.026 M$_{\sun}$ of $^{56}$Ni are present, if the lightcurve is solely powered by radioactive decay, although the $^{56}$Ni mass may be lower if CSM interaction contributes to the post-plateau luminosity. Prominent multi-peaked emission lines of H$\alpha$  and [O I] emerge after day 154, as a result of either an asymmetric explosion or asymmetries in the CSM.  The lack of narrow lines within the first two days of explosion in the likely presence of CSM interaction may be an example of close, dense, asymmetric CSM that is quickly enveloped by the spherical supernova ejecta.
\end{abstract}


\section{Introduction} \label{sec:intro}
Core-collapse supernovae (CCSNe) mark the death of stars more massive than $\sim$ 8 M$_{\odot}$.  Those stars that end their lives with portions of their hydrogen envelope remaining are classified as Type II events \citep[see ][for detailed reviews]{2017hsn..book..239A,2017hsn..book..195G,2017suex.book.....B}. Historically these events have been classified as Type II-P or Type II-L based on their lightcurve shapes.  Type II-P (``P" for plateau) show  a plateau phase of near constant luminosity in the lightcurve for $\sim$ 2--3 months after maximum light due to the long diffusion and recombination timescales of the hydrogen envelope, while Type II-L (``L" is for linear) shown an almost linear decline with no or short plateau phases. Recent work has suggested that this bi-modal classification is misleading, and in fact Type II SNe form a continuous class \citep{Anderson2014,Valenti16,2016AJ....151...33G}. Once the recombination phase ends, a sharp drop in luminosity occurs over a relatively short timescale, until the SN settles into the nebular phase where the lightcurve is powered primarily  by radioactive decay.   

Pre-explosion {\it Hubble Space Telescope (HST)} imaging  of Type II-P events point to red supergiant (RSG) stars as the most common progenitors \citep{VanDyk03,Smartt09,Smartt15}. RSGs do not form a homogeneous group, and variations in metallicity, initial mass, and mass-loss histories lead to diversity among the resultant SNe. Adopted mass-loss rates for RSGs generally range from $\sim$10$^{-6}$ to 10$^{-4}$ M$_{\sun}$ yr$^{-1}$, with average wind velocities of 10 km s$^{-1}$ \citep{MJ2011,Goldman17,Beasor18}. A recent study of early-time, high-cadence lightcurves in \citet{2018NatAs...2..808F} finds evidence for mass loss rates greater than 10$^{-4}$ M$_{\sun}$ in the majority of their RSG sample.  It is important to remember that these rates are for single star models, and since $\sim$75 $\%$ of massive stars in binaries have separations that can lead to interaction \citep{Kiminki12,Sana12,deMink14,2017ApJS..230...15M}, mass-loss rates and densities could vary if a companion is present.  

In $\sim$  8--9 $\%$ of CCSNe the circumstellar medium (CSM) surrounding the progenitor is photoionized or shock heated, creating narrow ($\sim$ 100 km s$^{-1}$) hydrogen emission lines in their spectra \citep{SmithSNPop11}. The narrow lines lend themselves to the name Type IIn, where the ``n" stands for narrow \citep{Schlegel1990}.  The progenitors of these IIn are likely special cases of evolved massive stars with pre-supernova outbursts, and could include RSGs, yellow hypergiants (YHGs), or luminous blue variables (LBVs) \citep{2007ApJ...656..372G,2009Natur.458..865G,SmithMassLoss14}. The SNe IIn 1998S \citep{Shivvers2015,Mauerhan12} and PTF11iqb \citep{PTF11iqb} are examples of objects that likely had RSG progenitors, and may have been classified as a normal Type II if they had not been observed so soon after explosion.

If a SN is observed early enough, before the SN ejecta overtake the surrounding material, narrow lines from slow CSM can be detected in otherwise normal SNe \citep{1985ApJ...289...52N,1994A&A...285L..13B,2000ApJ...536..239L,2007ApJ...666.1093Q}. If present, these early and brief spectral features can be used to infer properties about the progenitor star such as mass-loss history and composition \citep{GalYam2014,2014A&A...564A..30G,DaviesDessart19}. Additionally, if the CSM is dense enough, shock interaction with the SN ejecta can occur, converting the kinetic energy of the fast ejecta to radiative energy, thus increasing the luminosity of the SN. All of these features disappear within a week of explosion, eliminating them from the traditional class of Type IIn SNe.  To date, only a hand-full of objects have shown these early high ionization narrow emission lines including SN~2013cu \citep{GalYam2014}, SN~1998S \citep{Shivvers2015}, PTF11iqb \citep{PTF11iqb}, SN~2013fs \citep{Yaron2017}, and SN~2016bkv \citep{Hosseinzadeh18}. Others have shown a featureless, blue continuum with no lines \citep{Khazov16}. As we discuss below, SN~2017gmr was observed within 1.5 days of explosion, and showed no signs of narrow emission other than H$\alpha$ in early spectroscopy.

\begin{figure}
\includegraphics[width=3.3in]{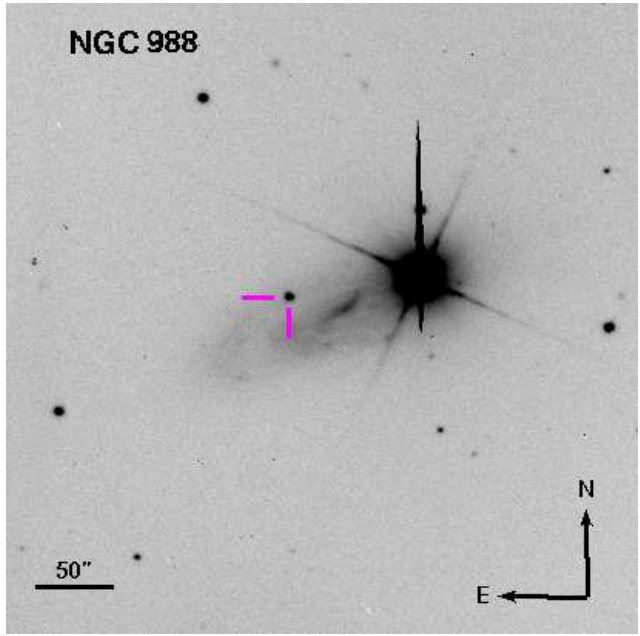}
\caption{SN~2017gmr in NGC~988 taken on 2017 November 25 in $V$-band with Super-LOTIS.  Image is 7$\arcmin$ $\times$ 7$\arcmin$.  }
\label{fig:image}
\end{figure}

SN~2017gmr was discovered at an RA(2000) $=$ 02$^h$35$^m$30$^s$.15, Dec(2000) $=-09\degr 21\arcmin 14\farcs 95$ during the course of the DLT40 one-day cadence SN search \citep[for a description of the survey, see][]{Tartaglia18} in the northeastern portion of NGC~988 (Figure \ref{fig:image}) on 2017 September 4.25 UT \citep[MJD~58000.266;][]{Valenti_17gmr}; it was given the designation DLT17cq by the DLT40 team, but we use the IAU naming convention and refer to it as SN~2017gmr throughout this work.  The discovery magnitude was $r$=15.12 ($M_r \approx -$16.3, given the distance modulus we adopt below), and DLT40 observations taken two days prior to discovery (MJD~57998.230) show no source at the position of the transient down to r$\gtrsim$19.4 mag ($M_r$$\gtrsim$$-$12.1), indicating the SN was caught very close to the time of explosion. In Section \ref{photometric} below we model the early-time light curves to constrain the explosion time and settle on MJD~57999.09 (2017 September 3.08) as the epoch of explosion, and adopt this value throughout the paper. 

Spectroscopic observations conducted on 2017 September 6.19 allowed classification of this object as a possible core collapse SN \citep{Pursimo_17gmr}; it was confirmed as a Type II with broad Balmer lines in emission and moderate reddening about 1 week after explosion, on 2017 September 10.2  \citep{Elias_17gmr}.  Adopting a redshift of z=0.00504 \citep{2004AJ....128...16K}, an  $H_{0}$ = 73.5 km s$^{-1}$ Mpc$^{-1}$ \citep{Riess2018}, and the Virgo infall velocity for the host NGC~988 given by NASA/IPAC Extragalactic Database (NED), $v_{Virgo}$ = 1438 $\pm$ 8 km s$^{-1}$, we obtain a $\mu$ = 31.46 $\pm$ 0.15 mag, or a distance of 19.6 $\pm$ 1.4 Mpc. NGC~988 is located in the same group as NGC~1084, the host galaxy of SN~2012ec, whose distance modulus was determined to be $\mu$ = 31.36 $\pm$ 0.15 mag in \citet{2019MNRAS.483.5459R}, bolstering our confidence in the assumed distance.  If we instead use the 3K CMB velocity $v_{CMB}$ = 1288 $\pm$ 16 km s$^{-1}$, or the Local Group velocity $v_{LG}$ = 1532 $\pm$ 5 km s$^{-1}$, this changes the distance to 17.5 $\pm$ 1.2 Mpc or 20.8 $\pm$ 1.5 respectively. The Virgo infall values are more consistent with our host galaxy line measurements, and fall nicely within other cosmological distance measurements so we will use that value throughout the paper.

The paper is structured as follows: in Section \ref{observations} observations and data reduction are outlined, the reddening estimation is presented in Section \ref{reddening}, in Section \ref{photometric} we discuss the optical and IR photometric evolution, Section \ref{spectroscopic} details the spectroscopic evolution of the object, in Section \ref{discussion} we lay out the implications of the observational data, and finally the results are summarized in Section~\ref{conclusion}.

\section{OBSERVATIONS} \label{observations}

A comprehensive optical and near-infrared (NIR) dataset has been collected on SN~2017gmr, with several major supernova collaborations contributing data.  These include the Las Cumbres Observatory's Global Supernova Project \citep[e.g.][]{Szalai19}, the NOT (Nordic Optical Telescope) Un-biased Transient Survey\footnote{http://csp2.lco.cl/not/} (NUTS), the  Public ESO Spectroscopic Survey for Transient Objects \citep[ePESSTO;][]{Smartt15}, and the Texas Supernova Spectroscopic Survey (TS$^{3}$).  Below we briefly list the instruments/telescopes used in obtaining data for SN~2017gmr but for ease of reading an accounting of reduction procedures is included in the Appendix.
 
Continued photometric monitoring of SN~2017gmr was done by the DLT40 survey's two discovery telescopes, the PROMPT5 0.4-m telescope at Cerro Tololo International Observatory and the PROMPT-MO 0.4-m telescope at Meckering Observatory in Australia, operated by the Skynet telescope network \citep{Reichart05}.  Additionally, an intense photometric campaign by the Las Cumbres Observatory telescope network \citep{Brown_2013}, under the auspices of the Global Supernova Project, was begun immediately after discovery, in the $UBVgri$ bands. Photometric data points were also taken at: 1) the 0.6-m Schmidt telescope at Konkoly Observatory in the $BVRI$ bands; 2) the 0.6-m Super-LOTIS telescope at Kitt Peak in the $BVRI$ bands; 3) the 2.0-m Liverpool Telescope and the Optical Wide Field camera (IO:O) in the $BVugriz$ bands; 4) the 2.56-m NOT Alhambra Faint Object Spectrograph and Camera (ALFOSC) in the $BVugriz$ bands; 5) the Asiago Schmidt 67/92-cm telescope in the $BVgri$ bands; 6) the 1.04-m Sampurnanand Telescope (ST) at Manora Peak, Nainital in $BVRI$ bands \citep{Sagar_1999}; (7) the 1.30-m Devasthal Fast Optical Telescope (DFOT) at Devasthal, Nainital in $UBVRIgri$ bands \citep{Sagar_etal_2012}; (8) the 2.01-m Himalayan Chandra Telescope (HCT) at Indian Astronomical Observatory (IAO) in Hanle, India \citep{Prabhu_etal_2010} in the $UBVRI$ bands; and (9) the 60-cm REM telescope in $griz$. Neil Gehrels Swift Observatory \citep[{\it Swift}]{GehrelsSwift} UV and optical imaging was obtained of the early portion of the light curve.  Furthermore, near-infrared (NIR) $J$, $H$, and $K_{s}$ images were taken with  NOTCam  on the 2.56-m NOT telescope and the REM 60-cm telescope.

Many optical spectra were taken with the robotic FLOYDS spectrographs on the 2-m Faulkes Telescope North and South \citep[FTN and FTS;][]{Brown_2013}. Other telescopes/instruments used were: 1) the Goodman spectrograph \citep{goodman} on the 4.1-m SOAR telescope; 2) the Intermediate Dispersion Spectrograph (IDS) on the 2.54-m Isaac Newton Telescope (INT); 3) the Inamori-Magellan Areal Camera \& Spectrograph \citep[IMACS;][]{imacs} on the 6.5-m Magellan Baade telescope; 4) the ALFOSC spectrograph on NOT; 5) the ESO Faint Object Spectrograph and Camera 2 (EFOSC2) on the 3.58-m New Technology Telescope (NTT), 6) the Beijing Faint Object Spectrograph and Camera (BFOSC) on the Xinglong 2.16m telescope; 7) the Asiago Faint Object Spectrograph and Camera (AFOSC) on the Asiago 1.82-m telescope; 8) the FOcal Reducer and low dispersion Spectrograph 2 \citep[FORS2;][]{fors} on the 8.2-m Very Large Telescope (VLT); 9) the Himalaya Faint Object Spectrograph and Camera (HFOSC) on HCT; 10) the Boller \& Chivens (B\&C) Spectrograph mounted on the Asiago 1.22-m telescope; 11) the Low Resolution Spectrograph \citep[LRS2;][]{Chonis16} on the effective 10-m Hobby-Eberly Telescope (HET); and  12) the Boller \& Chivens (B\&C) Spectrograph mounted on the 2.3-m Bok telescope on Kitt Peak. Further, a moderate-resolution spectrum was  obtained with the Blue Channel (BC) spectrograph on the 6.5-m MMT. High-resolution echelle spectra were taken with the  HIgh-Resolution Echelle Spectrograph  \citep[HIRES;][]{HIRES} on Keck and the Magellan Inamori Kyocera Echelle instrument \citep[MIKE;][]{MagellanMike} on the Magellan Clay telescope. NIR spectra were taken with the Gemini Near-Infrared Spectrograph (GNIRS) at Gemini North Observatory \citep{Elias06}, the Folded-port InfraRed Echellette \citep[FIRE;][]{Simcoe13} on Magellan Baade, SpeX \citep{Rayner03} on the NASA Infrared Telescope Facility (IRTF), and the Son OF ISAAC (SOFI) spectrograph mounted on the NTT \citep{SOFI}.

 \begin{figure}
\includegraphics[width=3.5in]{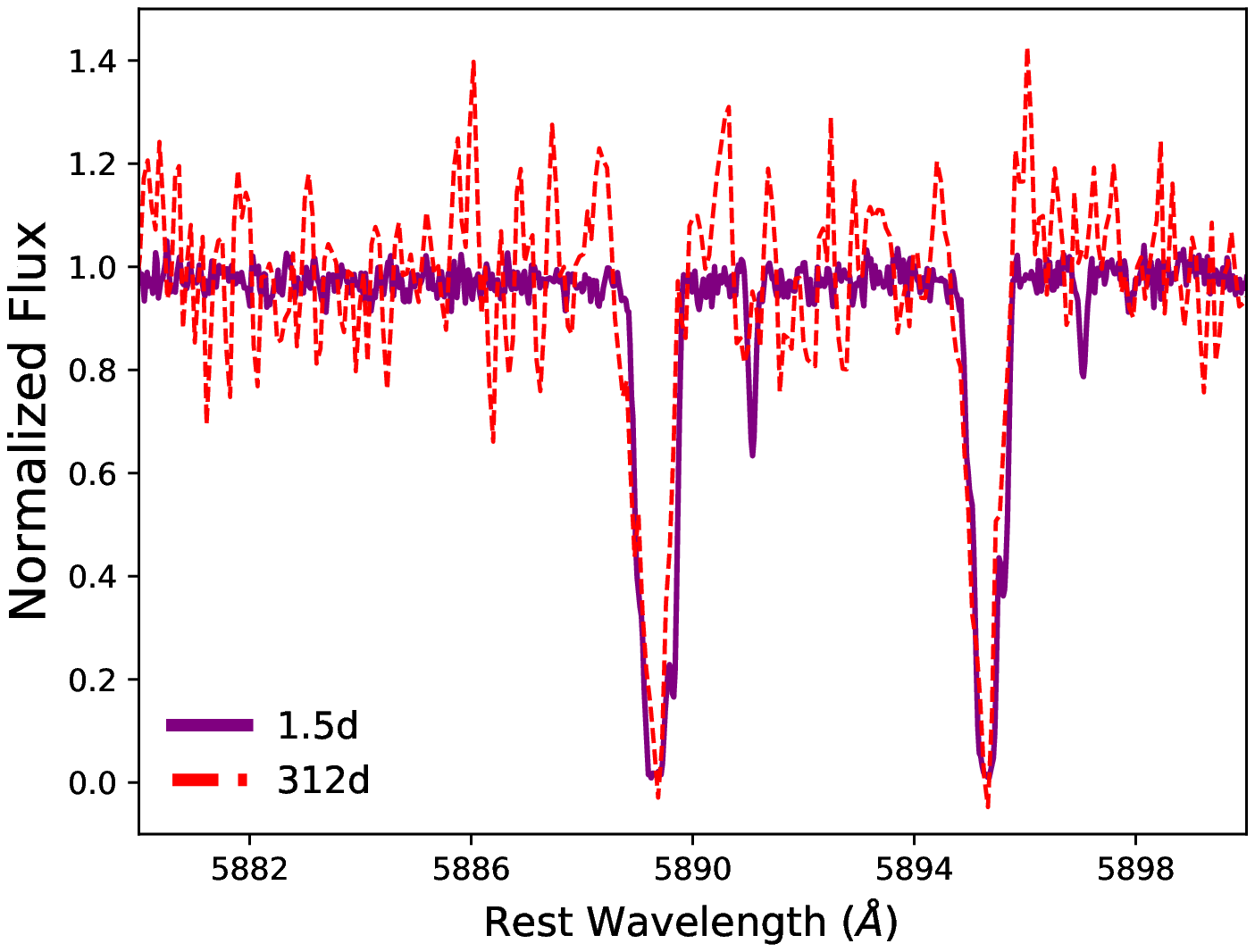}
\includegraphics[width=3.5in]{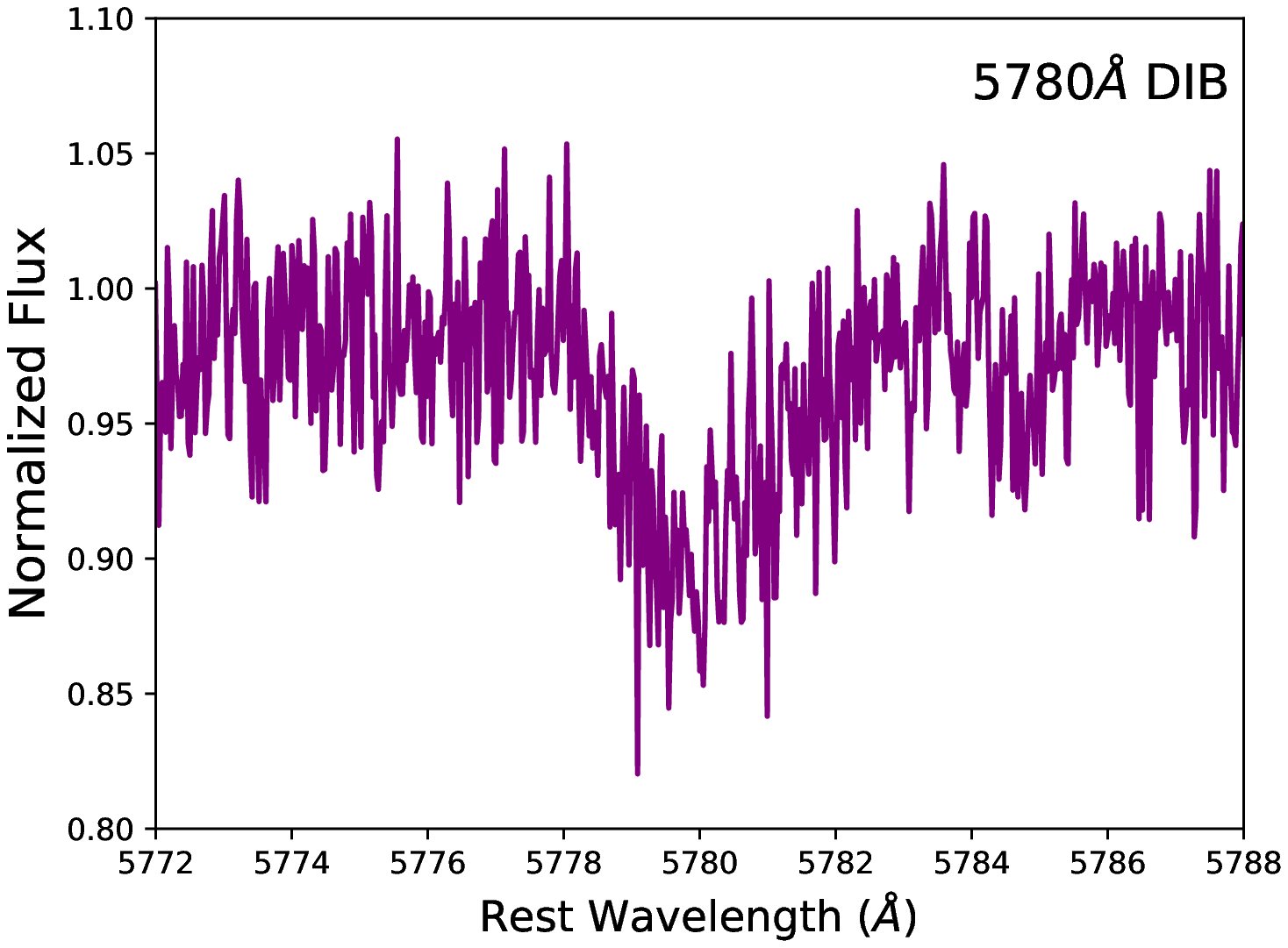}
\caption{Keck HIRES spectra (purple) from day 1.5 showing the region around the NaID lines (top) and the $\lambda$5780 DIB feature (bottom). The red NaID spectra is from Magellan/MIKE echelle spectra on day 312.  }
\label{fig:NaID}
\end{figure}

\begin{figure}
\includegraphics[width=3.5in]{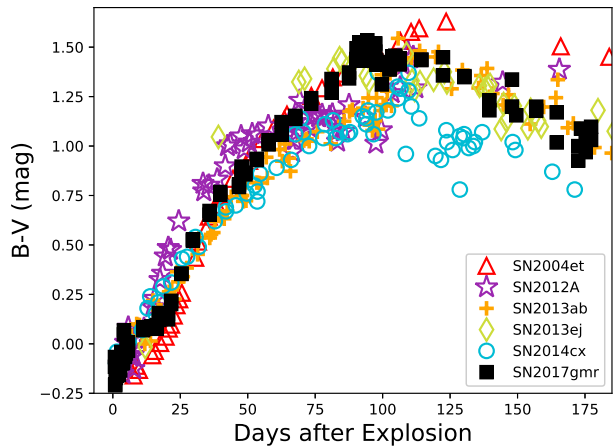}
\caption{$B-V$ color evolution of SN~2017gmr (black) compared with other Type II-P SNe from the literature. All data have been corrected for reddening as indicated from the corresponding references.  The data come from sources listed in Section \ref{reddening}.   }
\label{fig:colorevo}
\end{figure}

\section{Reddening Estimation} \label{reddening}
  The Milky Way line-of-sight reddening for NGC~988 is $E(B-V)_{MW} = 0.024$ mag \citep{2011ApJ...737..103S}. \citet{Elias_17gmr} noted strong host Na ID absorption with an equivalent width (EW) of 1.45 \AA\ on day 6, resulting in an estimation of a total $E(B-V)_{tot} = 0.23$ using the relation presented in \citet{2003LNP...598...21T}.  From the high-resolution Keck HIRES spectrum taken $\sim$ 6 hours after discovery (Figure \ref{fig:NaID}, top) we measure EWs of the individual Na ID lines of 0.75 and 0.62 \AA\, similar to the combined value found by \citet{Elias_17gmr}.  Unfortunately, the relationship between Na ID EW and dust extinction presented in \citet{2012MNRAS.426.1465P} saturates around 0.2 \AA\, requiring alternative methods for the reddening estimation of SN~2017gmr.

From the same early high-resolution spectrum we also detect the 5780 \AA\ diffuse interstellar band (DIB) absorption feature ((Figure \ref{fig:NaID}, bottom), which can be used to estimate the extinction $A_{V}$ \citep{Phillips13}. We obtain an EW of 0.22 \AA\, which corresponds to $A_{V} = 1.14$ mag, or an $E(B-V)_{tot} = 0.36$ mag using an $R_{V}$ = 3.1 and the reddening law of CCM \citep{1989ApJ...345..245C}.  Note that the uncertainty from this relationship is limited to $\pm$50$\%$, which only constrains the extinction to between $A_{V}$ $\approx$ 0.6-1.7 mag.    
  
We also compare the $B-V$ color of SN~2017gmr during the plateau phase to other Type II SNe with published reddening estimates and adjust the $E(B-V)$ accordingly until we have a similar fit (similarly to that done by \citealt{Tartaglia18}). Comparison with SNe 2004et \citep{2006MNRAS.372.1315S}, 2012A \citep{Tomasella2013}, 2013ab \citep{SN2013ab}, 2013ej \citep{Bose2015}, and 2014cx \citep{SN2014cx}, shown in Figure \ref{fig:colorevo},  constrain the reddening to  $E(B-V)$ = 0.30 $\pm$0.1 mag. 

\begin{figure*}
\includegraphics[width=7.1in]{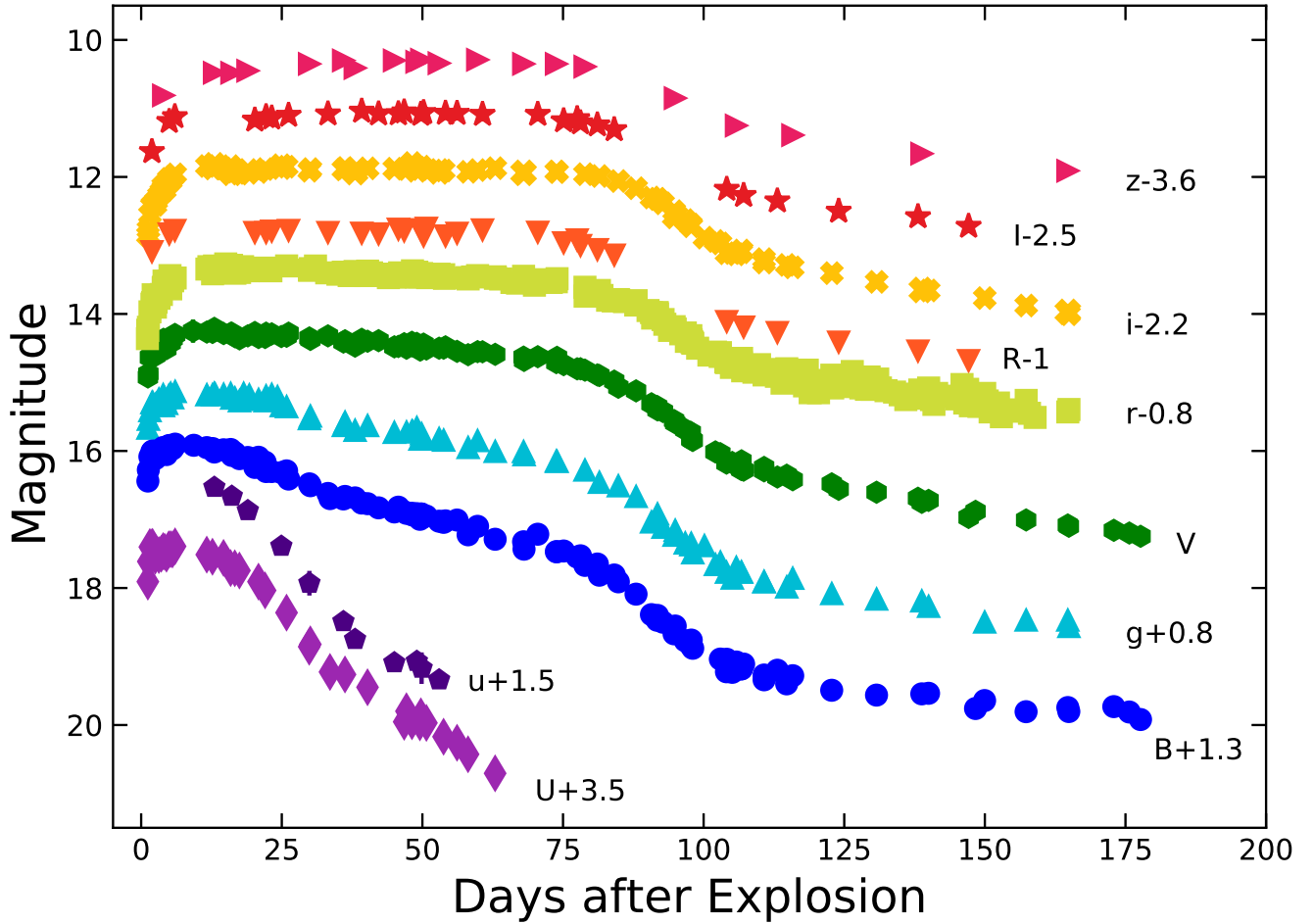}
\caption{Optical photometry of SN~2017gmr, shifted by constants for ease of viewing. Marker size is larger than uncertanties.  The dataset is tabulated in Table~\ref{tab:optphot}. The adopted date of explosion is considered to be MJD 57999.09 (2017 September 3.1 UT) as described Section \ref{modeling}.}
\label{fig:fulllc}
\end{figure*}

As another constraint we have compared our unreddened spectra with optical spectra of SN~2004et, a prototypical Type II-P, from similar epochs and applied reddening corrections until the spectra had a matching continuum slope.  SN~2004et has a measured $E(B-V) = 0.43$ mag \citep{2006MNRAS.372.1315S}, and comparisons on both day 7 and day 84 yield a total $E(B-V)= 0.30$ mag in SN~2017gmr.  As this value is consistent with the other two estimates we settle on a value of $E(B-V) = 0.30$ mag as our final reddening estimation, with the caveat that there may be somewhat large uncertainties.  This is the standard value that will be used throughout the paper.

\section{Photometric Evolution} \label{photometric}

\subsection {Optical Lightcurve}
\begin{figure*}
\includegraphics[width=7.1in]{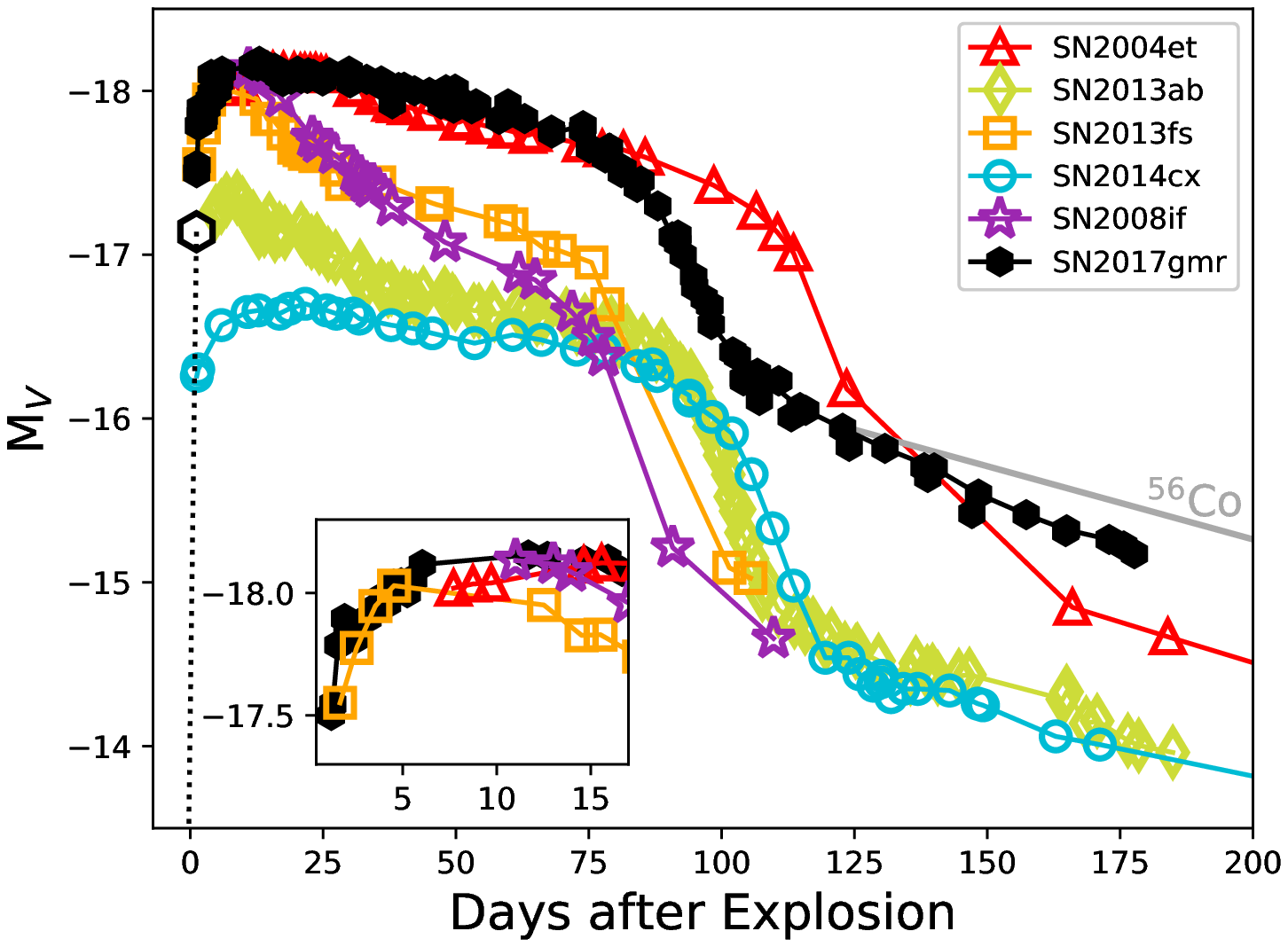}
\caption{Absolute $V$-band lightcurves of a sample of Type II SNe.  The inset in the lower left shows the comparison over the first 15 days among SNe 2017gmr, 2004et, 2013fs, and 2008if. Data are from \citet[SN~2004et]{2006MNRAS.372.1315S},  \citet[SN~20013ab]{SN2013ab}, \citet[SN~2014cx]{SN2014cx}, \citet[SN~2008if]{Gutierrez2017a}, and \citet[SN~2013fs]{Valenti16}. }
\label{fig:lccomp}
\end{figure*}

The full optical lightcurve can be seen in Figure \ref{fig:fulllc}, and the $V$-band lightcurve compared to other Type II SNe is shown in Figure \ref{fig:lccomp}. For reference, the $r$-band discovery magnitude is shown as an open hexagon while the dotted line connects the pre-explosion upper-limit $r$-band magnitude two days prior in Figure \ref{fig:lccomp}.    Overall the shape is that of a typical Type II supernova with an extended plateau, albeit on the brighter end with a maximum $M_V$ = -18.3 mag.  The maximum occurs at $\sim$ 6 days after explosion for the $U$ and $B$ bands, $\sim$ 8 days for $g$ and $V$, and closer to 10 days for $r$ and $i$. This is consistent with the average rise times seen for the majority of Type II SNe \citep{2015MNRAS.451.2212G, 2016ApJ...820...33R, 2018NatAs...2..808F}.

The lightcurves then remain at a relatively constant magnitude for the next 75 days until the fall off the plateau begins around day 85, with decline rates of 0.027, 0.011, and 0.003 mag day$^{-1}$ in $B$, $V$, and $i$ respectively.   Using the method described in \citet{Valenti16}, we obtain the point at half of the fall at MJD 58093.5 $\pm$0.4, or 95 days after our estimated explosion date. Between day 85 and 105 the V-band lightcurve drops by 1.5 mag. This moderate post-plateau drop is on the lower end but consistent with other II-P SNe, particularly higher luminosity events \citep{Valenti14}.  

 The plateau length of SN~2017gmr is on the shorter side for comparable objects and has an average $M_{V}$ = $-$17.8 mag (Figure~\ref{fig:lccomp}), a value noticeably brighter than the norm (but similar to SN~2004et). According to \citet{Anderson2014}, \citet{Faran2014}, and \citet{2016AJ....151...33G}, more luminous Type II-P SNe tend to exhibit shorter plateau durations, which coincides with the overall picture of SN~2017gmr.  SN~2017gmr, SN~2013fs, SN~2004et, and SN~2008if  all show similar luminosities and evolution over the first few days (Figure~\ref{fig:lccomp} inset), but then evolve to drastically different lightcurve shapes.  While SN~2017gmr and SN~2004et change very little over the first 3 months, SN~2013fs and SN~2008if show evolution more akin to Type IIL SNe, with a larger drop in luminosity over the first $\sim$ 75 days.  

\subsection{The early U-bump}\label{sec:ubump}
\begin{figure}
\includegraphics[width=3.5in]{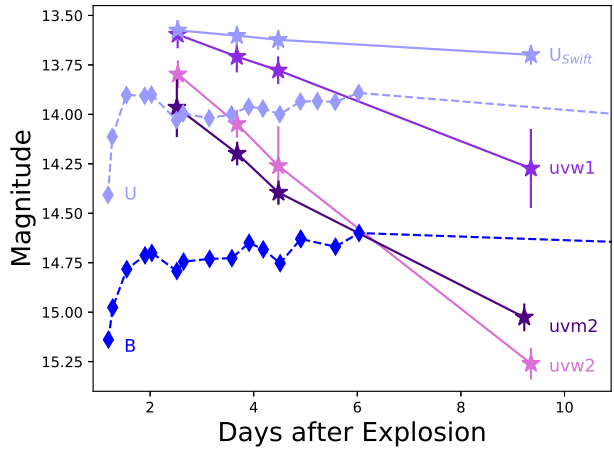}
\caption{{\it Swift} photometry of SN~2017gmr compared with ground-based $U$ and $B$-band photometry from Las Cumbres Observatory. The {\it Swift} photometry is tabulated in Table~\ref{tab:swiftphot}. }
\label{fig:Ulc}
\end{figure}
One rather intriguing feature seen in the early lightcurve of SN~2017gmr is the bump in luminosity that occurs a couple of days post-explosion, particularly in the bluest bands.  In Figure \ref{fig:Ulc} we show the ground-based $U$ and $B$ observations along with the {\it Swift} UV.  From the $U$ and $B$ data we see a sharp rise over the first 2 days, then a drop of roughly 0.2 mag and 0.1 mag in $U$ and $B$ respectively, then a slow rise over the next few days back to the peak value. Unfortunately, no {\it Swift} data exists prior to day 2 so the lightcurve behavior in the UV bands is unknown over the same time period.  It is also possible that we are seeing undulations in the $U$ and $B$ lightcurves due to inhomeganities in the CSM, particularly in some cases where the magnitude changes are larger than the uncertainties.

Models recently produced by \citet{Moriya18} do show this small bump in luminosity in the $u$ and $g$ bands with certain mass-loss and density configurations (see also \citealt{Morozova18}).  The key to creating this early bump is to have moderately dense CSM close to the progenitor. The Type II-P SN~2016X showed a similar bump  in the {\it Swift} $UV$ lightcurve over the first few days after explosion, although it did not seem to be present in the optical bands \citep{SN2016X}.  Their explanation for the initial lightcurve peak was a shock breakout cooling effect, but as we discuss in Section \ref{modeling}, we cannot fit this bump with standard shock-cooling models.  

\subsection{Late Time Lightcurve}
As we show in Figure \ref{fig:lccomp}, the radioactive tail of SN~2017gmr does not show the exponential decline of $^{56}$Co decay of 0.98 mag 100 d$^{-1}$ \citep{1989ApJ...346..395W}. While the $B$-band declines around 0.9 mag 100 d$^{-1}$, $V$ and $i$ decline by 1.5 and 1.4 mag 100 d$^{-1}$, respectively. By our last photometric observations around day 175, the $V$-band lightcurve is about 0.5 mag fainter than expected.  The same behavior is seen in the bolometric lightcurve, as we discuss below. The deviation from predicted $^{56}$Co decay can be explained by incomplete gamma ray trapping, a decrease in the energy input from  shock interaction, as dust production in the ejecta, or some combination of the three.  

Incomplete gamma-ray trapping has been documented in other Type II-P SNe. \citet{Anderson2014} found that the more luminous the SN, the greater the deviation from the expected decay rate and attributed it to low ejecta mass. Highly energetic explosions can also have large expansion velocities, which in turn leads to weaker trapping.  Alternatively, if the distribution of $^{56}$Ni is very asymmetric or mixed in the ejecta, the escape probability could be greater. If CSM interaction is occurring it can also contribute to the luminosity at late times and would not follow the predicted rate of $^{56}$Co decay.  We will discuss these possible scenarios further in Section \ref{discussion}.

\begin{figure}
\includegraphics[width=3.5in]{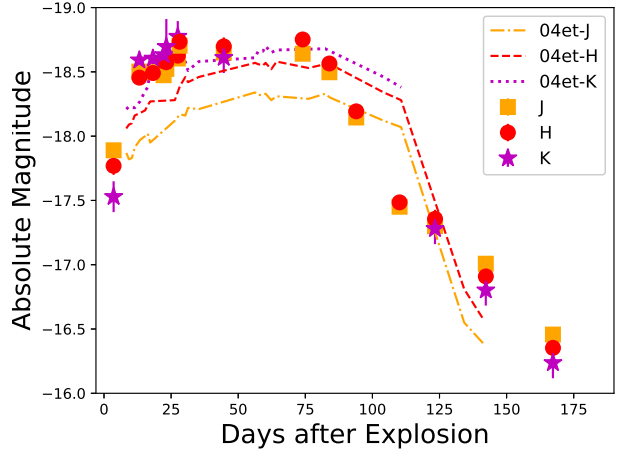}
\caption{NIR lightcurve of SN~2017gmr in absolute magnitudes using $E(B-V)$=0.30 and $\mu$=31.46 mag.  Also shown for comparison is the NIR photometry of SN~2004et from \citet{Maguire10} corrected for an $E(B-V)$ = 0.41 and a $\mu$=29.4 mag \citep{NGC6946TGRB}.   }
\label{fig:irlc}
\end{figure}

\subsection{Infrared Lightcurve}

Multiple epochs of NIR data were obtained over the first 160 days of evolution.  The NIR luminosity rose over the first 30-40 days after explosion (Figure \ref{fig:irlc}). This was followed by a few weeks of nearly constant luminosity, then starting around day 75 a steady decline begins in all filters  and continues until our last observed epoch.  We have plotted the NIR lightcurves of SN~2004et from \citet{Maguire10} as a comparison, and it indicates that the NIR plateau is much shorter for SN~2017gmr than SN~2004et, and that likely the late-time NIR luminosity is greater for SN~2017gmr as well.

\subsection{Color Evolution}

The $B-V$ color evolution of SN~2017gmr and a comparison to other SNe are shown in Figure \ref{fig:colorevo}.  As in other Type II SNe, the color is initially blue and evolves rapidly towards the red as the large envelope of the RSG progenitor expands and cools, until it reaches the recombination phase and the rate slows \citep{dejaegar18}.  This continues over the duration of the optically thick plateau phase until a peak value of $B-V = 1.5$ mag. After day 100, once the exponential decline phase begins, the color gradually becomes bluer again.

 \subsection{Bolometric Lightcurve}
The abundance of photometric data has allowed us to straightforwardly create a quasi-bolometric lightcurve using the routine \textsc{superbol} \citep{supbol18}. Following the description in \citet{Nicholl16}, the reddening and redshift corrected photometry in each band was interpolated with the $g$-band as reference, then converted to a spectral luminosity (L$_{\lambda}$).  The bolometric luminosity was then computed from the integration of the  SED for each epoch.

\begin{figure}
\includegraphics[width=3.1in]{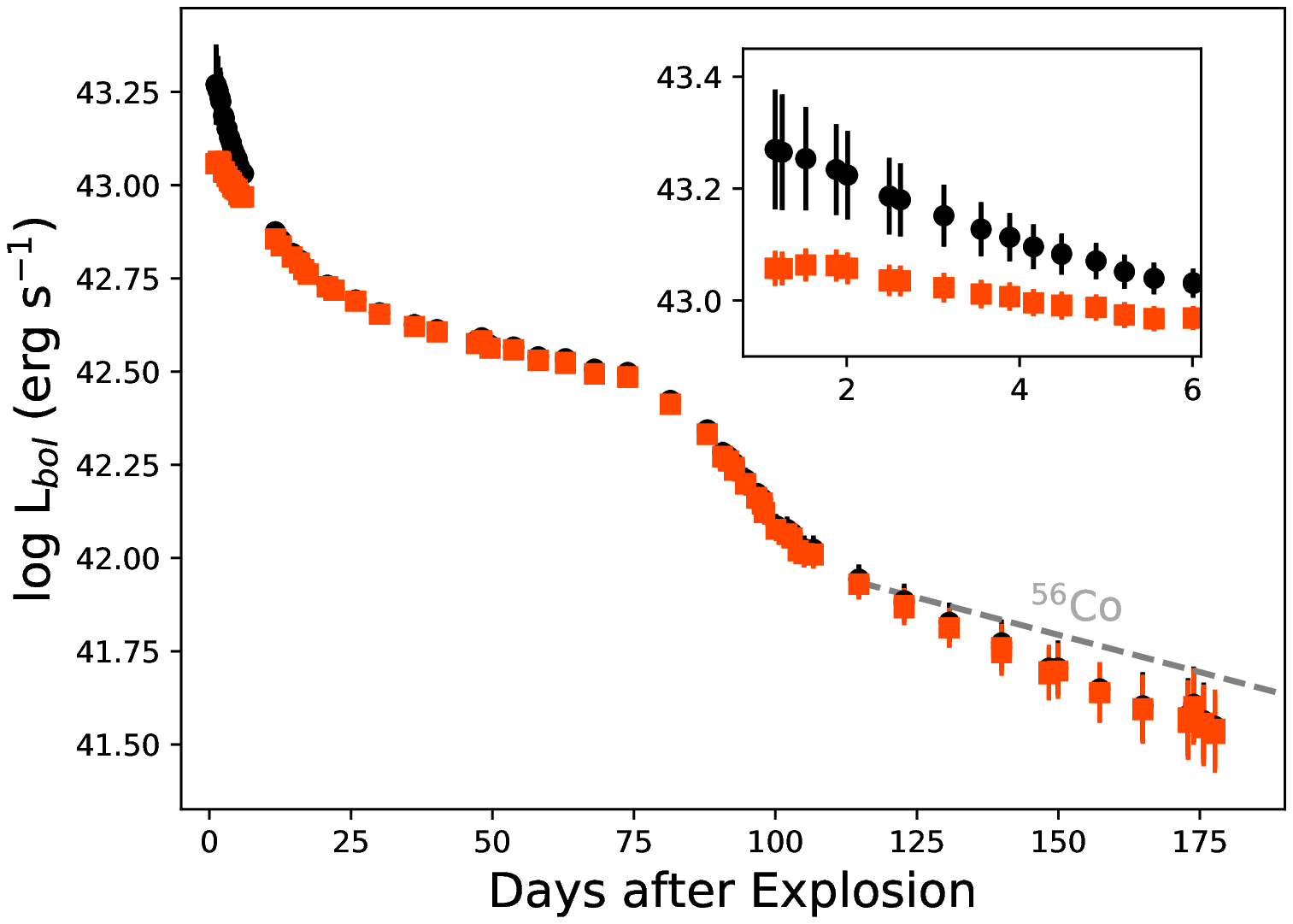}
\includegraphics[width=3.3in]{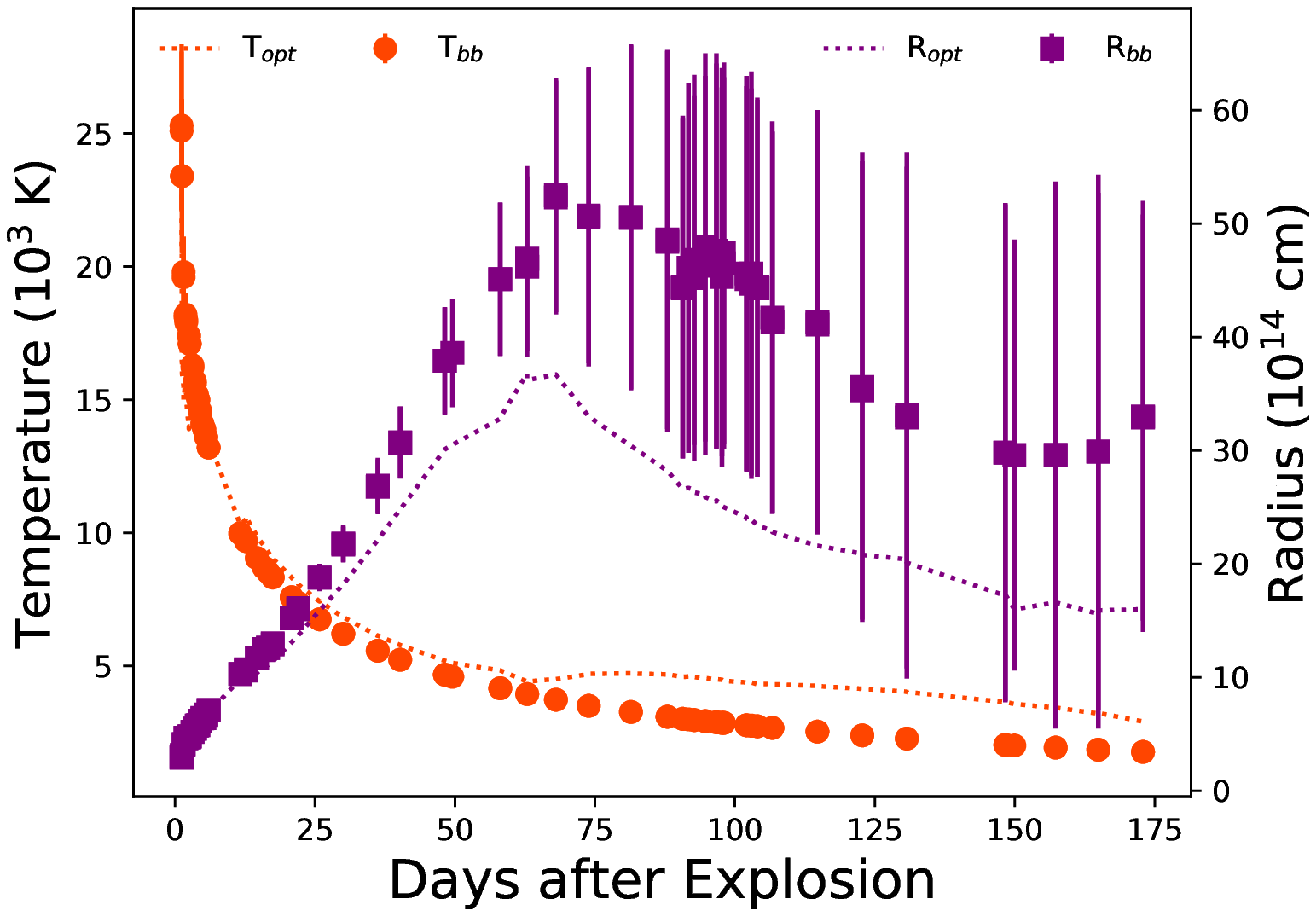}
\caption{Top: Bolometric lightcurve integrated from  NUV to NIR.  Inset shows a zoom in of the first few days. The red points indicate the observed luminosity, while the black points come from blackbody corrections to the data. The $^{56}$Co decay rate is indicated in gray.  Bottom: Temperature and radius evolution of SN~2017gmr derived from the photometry. The temperature is plotted in red, and the radius in purple.}
\label{fig:bolo}
\end{figure}

In Figure \ref{fig:bolo} we show the bolometric lightcurve produced from the observations (red), and those obtained with blackbody corrections (black), as well as the bolometric temperature ($T_{bol}$) and bolometric radius ($R_{bol}$) shown in the bottom of Figure \ref{fig:bolo}. The red lightcurve is pseudo-bolometric, and is constructed by integrating under the filters from UV to IR. $Swift$-UV coverage does not extend past $\sim$ 9d, so a first-order polynomial is fit to the data and extended out to later epochs.  As the contribution to the total bolometric luminosity falls quickly after the first few weeks this does not add much uncertainty.  The data have been corrected for an $E(B-V) = 0.30$ mag and adopting the distance modulus $\mu$ = 31.46 mag. 

As we mention above, the late-time lightcurve falls faster than expected for a fully-trapped $^{56}$Co decay, with L$_{bol}$ roughly 5 $\times$ 10$^{41}$ ergs s$^{-1}$ fainter that predicted  on day 165.  Integrating over the entire bolometric lightcurve gives a total radiated energy of 3.5 $\times$ 10$^{49}$ ergs in the first 175 days.

\subsection{A Search for pre-SN Outbursts}

With the advent of high cadence transient searches in the last decade, several instances of pre-SN outbursts have been observed directly in the months to years before explosion \citep[e.g.][]{2013ApJ...779L...8F,Mauerhan13,Ofek13,Ofek14,Elias16,2016MNRAS.459.1039T,2019MNRAS.482.2750R}, although overall detectable outbursts are rare \citep{2015MNRAS.450..246B,2015ApJ...811..117S}. These outbursts are generally associated with SNe that have substantial circumstellar material as evidenced by their SN IIn-like behavior.   However, many standard Type II-P/L SNe also show evidence for CSM material either as narrow emission lines in their early time spectra \citep[e.g.][]{Khazov16} or early peaks in their light curves \citep{Morozova17}.  This CSM could have been deposited in the years or decades prior to explosion, and could have been accompanied by faint pre-SN outbursts, as has recently been suggested in the gravity wave driven scenario of \citet{2014ApJ...780...96S,Fuller17}, or in unsteady nuclear burning events or binary interaction \citep{2014ApJ...785...82S}.

The field of NGC~988 was observed by the DLT40 survey 56 times between January 2015 and September 2017, just prior to the explosion of SN~2017gmr.  During much of this time period the DLT40 survey was coming online, with some prolonged down periods.  No precursor outbursts were observed down to a typical limiting magnitude of $r$$\sim$19--19.5 mag ($-$12 $>$M$_r$$>$$-$12.5). We can therefore rule out bright eruptions like SN imposters or LBV eruptions with roughly $M_{r}$ = $-$14 mag lasting several months, but not fainter or short-lived outbursts.  This includes those LBV eruptions that have been found to have magnitudes of only $M_{r}$ = $-$10 or $-$11 mag \citep{2011MNRAS.415..773S}.

\begin{figure*}
\includegraphics[width=6.5in]{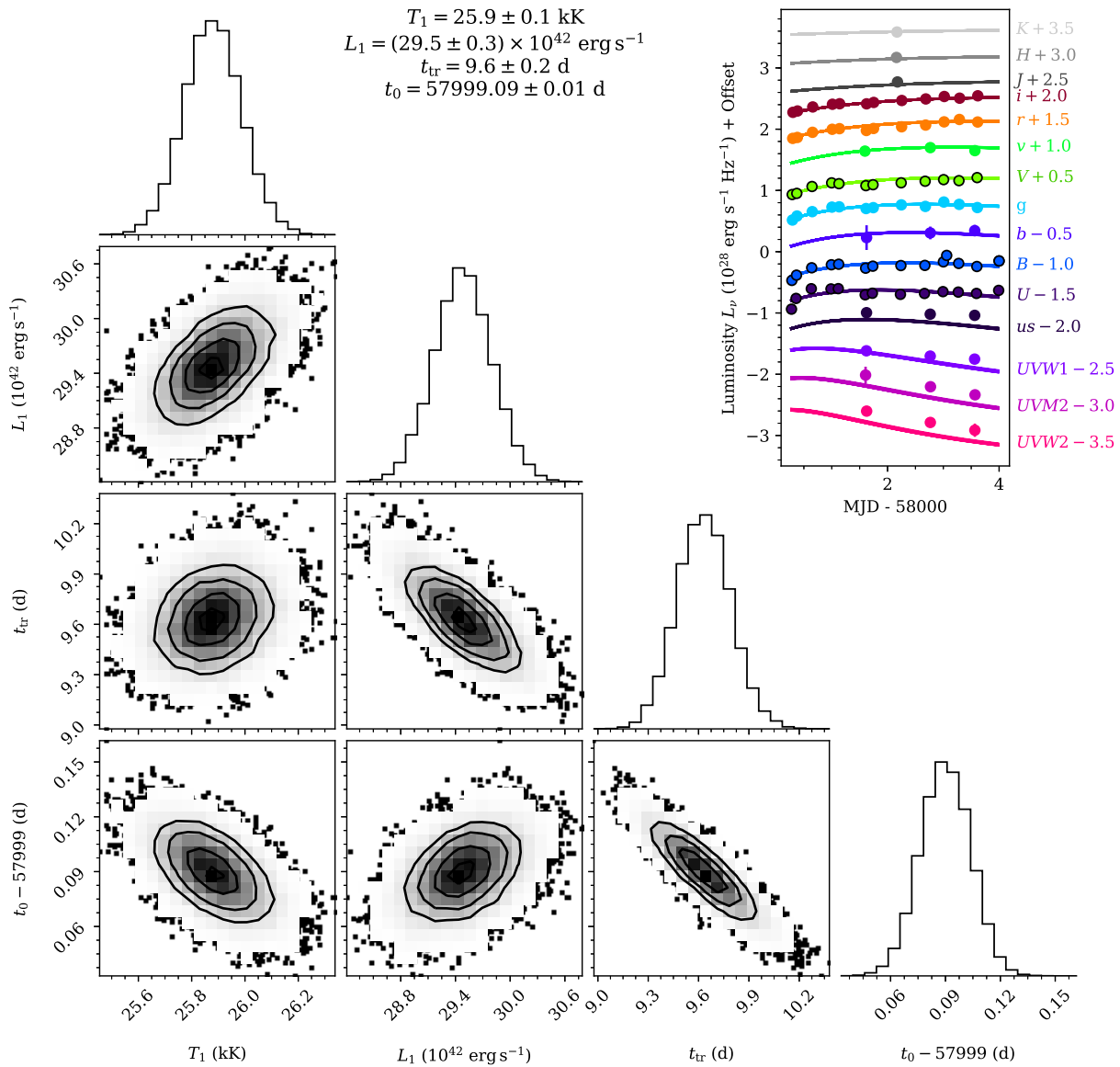}
\caption{Posterior probability distributions of various parameters of SN~2017gmr calculated using the methods described in \citet{Hosseinzadeh18}, who applied them to the SN~2016bkv.  We show the temperature and luminosity 1 day after explosion ($T_{1}$, $L_{1}$), the time of explosion ($t_{0}$), and the time to envelope transparency ($t_{tr}$). The top right panel shows 100 fits randomly drawn from the MCMC routine \citep{griffin_hosseinzadeh_2019_2639464} fit to the photometry.  The fits appear as a thick solid line due to the goodness of fit. Deviations between the lightcurve points and the fits are likely due to early CSM interaction. }
\label{fig:sc2}
\end{figure*}

\subsection{Early Lightcurve Modeling} \label{modeling}

Due to the well-sampled photometric data over the first few days after explosion in SN~2017gmr, we are able to model the early time lightcurves using the prescriptions outlined in \citet{SapandWax17}.  To do this we employed the code presented in \citet{griffin_hosseinzadeh_2019_2639464} and described in \citet{Hosseinzadeh18}, which uses a MCMC routine to fit the lightcurve in each photometric band and outputs posterior probability distributions for physical parameters, such as the time of explosion, the temperature, luminosity, and radius one day after explosion, and the time at which the envelope becomes transparent. Data was only fit up to day 4 to still lie within the validity range described by \citet{2017ApJ...848....8R}. The best fits to our data are shown in Figure \ref{fig:sc2}.  

One day after explosion the modeled temperature is 25.9 $\pm$ 0.1 $\times$ 10$^{3}$ K (kK) with a radius of 489 $\pm$ 22 R$_{\sun}$ (3.4 $\times$ 10$^{13}$ cm) and a luminosity of 2.9 $\pm$ 0.03 $\times$ 10$^{43}$ erg s$^{-1}$. The estimated progenitor radius is on the small end for a RSG which theoretically can range in size from $\sim$ 100 -- 1500 R$_{\sun}$ \citep{2017ars..book.....L}, but is commensurate with observations of some Galactic RSGs \citep[for example]{2018A&A...614A..12M,2017A&A...597A...9W}. From these fits we also derive an explosion date of MJD 57999.09 $\pm$ 0.01 d. This value is further bolstered by our first observation obtained on MJD 58000.27, or just over a day after the estimated explosion date, and our last non-detection on MJD 57998.22. This is also consistent with the UV photometry obtained 2.5 days after discovery which does not show a rise to peak that is seen in other bands (Figure \ref{fig:Ulc}).

\begin{figure*}
\includegraphics[width=7.1in]{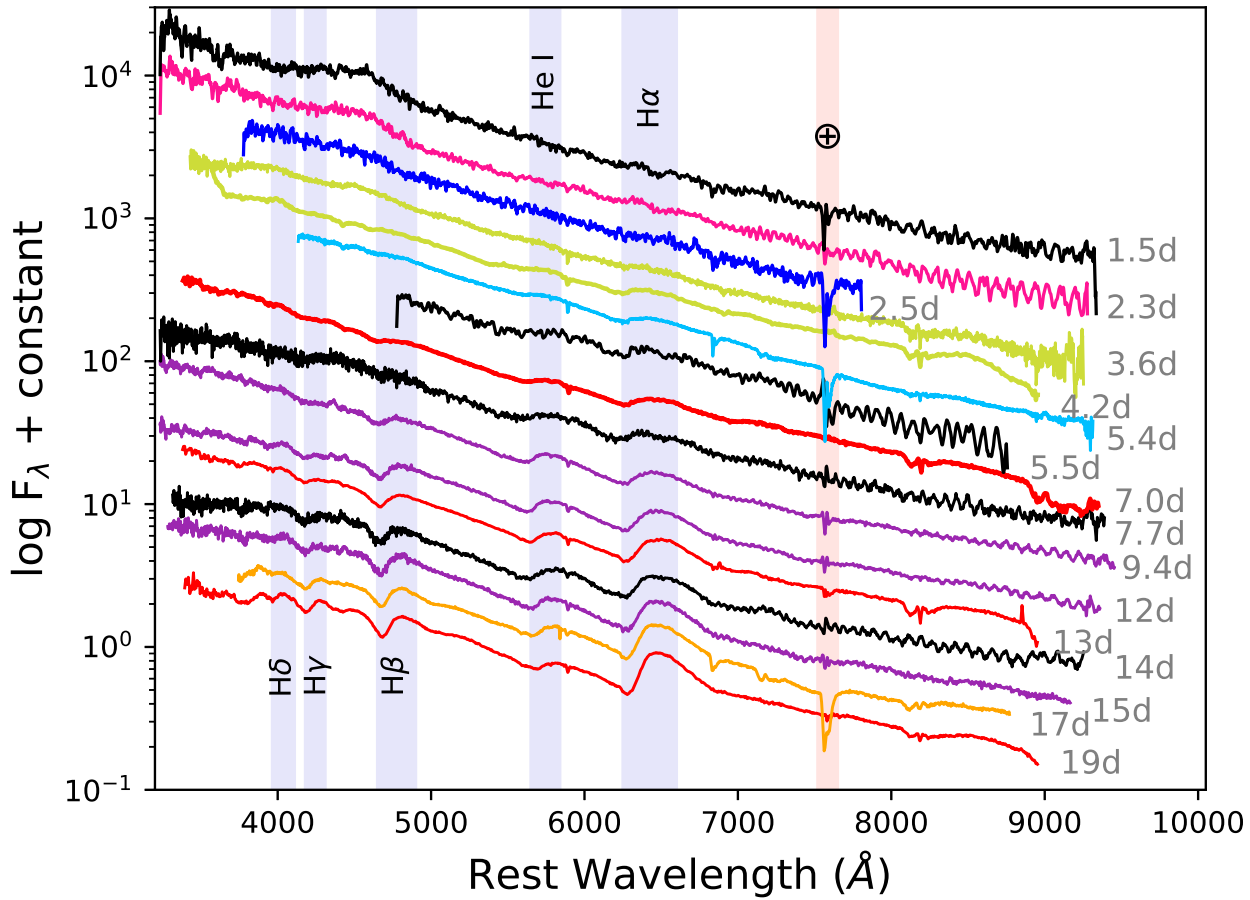}
\caption{Optical spectral sequence of SN~2017gmr up until 19 days after explosion. The color of each spectrum represents a particular instrument$+$telescope pair that corresponds to the same post-explosion date as listed in the optical spectroscopy log presented in Table~\ref{tab:optspec}. }
\label{fig:earlyspec}
\end{figure*}

\section {Spectroscopic Evolution} \label{spectroscopic}

\begin{figure*}
\includegraphics[width=7.1in]{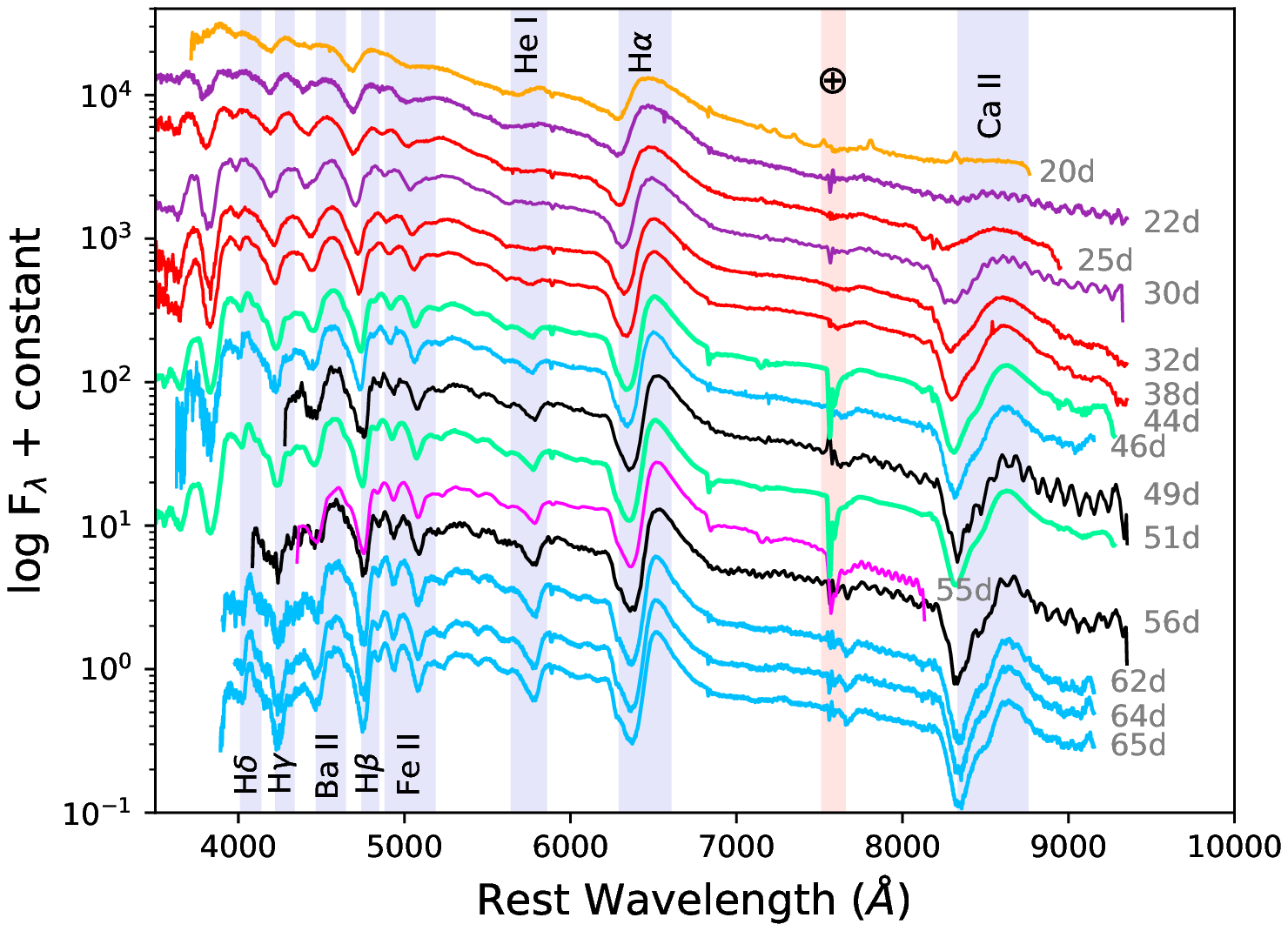}
\caption{Same as Figure \ref{fig:earlyspec} but for 20 to 65 days after explosion. }
\label{fig:midspec}
\end{figure*}

\begin{figure*}
\includegraphics[width=7.1in]{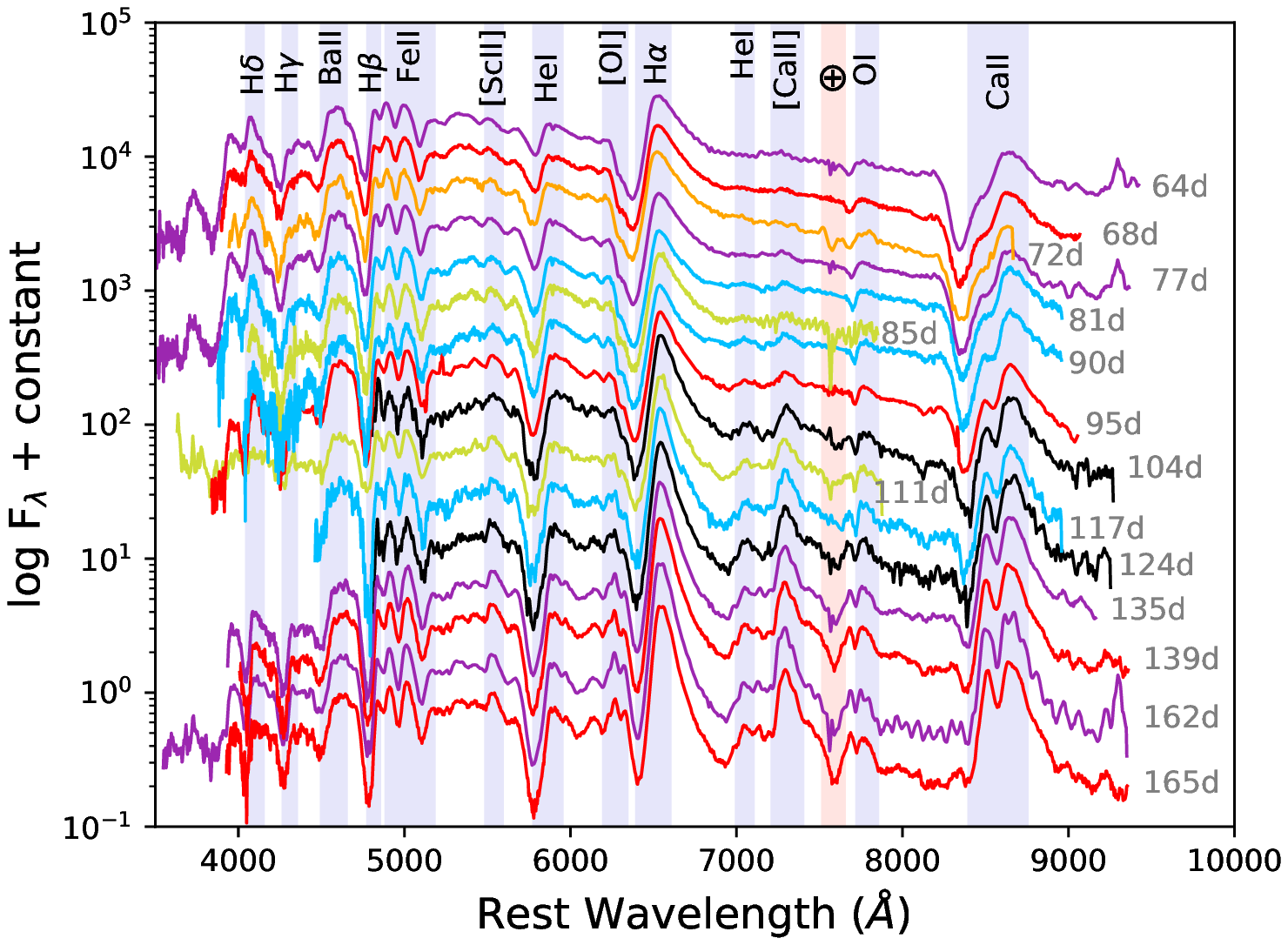}
\caption{Same as Figure \ref{fig:earlyspec} but for 64 to 165 days after explosion.  }
\label{fig:latespec}
\end{figure*}

\subsection{Optical Spectra}

The early spectra, shown in Figure \ref{fig:earlyspec}, are typical for a young II-P supernova, displaying a blue, mostly featureless continuum. Only strong interstellar NaID absorption lines, and a broad emission feature around 4600 \AA\ (likely \ion{He}{2} $\lambda$4686) are seen. Neither the low-resolution FLOYDS spectrum or the high-resolution Keck spectrum, taken within hours of discovery,  show signs of narrow high-ionization lines, other than narrow  55 km s$^{-1}$ H$\alpha$ seen in the Keck HIRES echelle spectrum (inset \ref{fig:earlyspec}). This is different from other early-detected CCSNe which can show features of highly ionized nitrogen and carbon along with He and H. This is discussed further in Section \ref{flash}.

As the photosphere begins to cool, the continuum becomes redder and broad Balmer emission lines begin to appear with P-Cygni absorption features.  When H$\alpha$ becomes pronounced a week after explosion the peak appears blueshifted, centered at $-$5000 km s$^{-1}$.  This is a common occurrence in Type II-P SNe where the opaque hydrogen envelope preferentially obscures the redshifted, receding side of the line \citep{2005A&A...437..667D,Anderson2014}.  As the recombination front moves through the envelope, the red side becomes visible again and the emission line peak becomes more symmetric. 

Around a month after explosion, the SN is well into the plateau phase and the \ion{Ca}{2} IR triplet centered around 8600 \AA\ emerges, along with a forest of metal lines blueward of 5000 \AA\ (Figure \ref{fig:midspec}).  In particular, lines of \ion{Fe}{2}, including \ion{Fe}{2} $\lambda$4924, $\lambda$5018, and $\lambda$5169  can be seen.

By the end of the plateau phase other broad lines such as \ion{Ba}{2} $\lambda$6142, [\ion{Sc}{2}] $\lambda$5527, $\lambda$5658, and $\lambda$6246 (blended with [\ion{O}{1}]), 
and [\ion{O}{1}] $\lambda\lambda$6300,6364 appear in the nebular spectra (Figure \ref{fig:latespec}). Redward of H$\alpha$, strong [\ion{Ca}{2}] $\lambda\lambda$7291,7324 is seen, flanked on either side by \ion{He}{1} $\lambda$7065, \ion{Fe}{2} $\lambda$7155 and \ion{O}{1} $\lambda$7774.  What appears to be \ion{K}{1} $\lambda\lambda$ 7665,7699 is also detectable and distinct from \ion{O}{1} by $\sim$ day 120. The emergence of the \ion{He}{1} $\lambda$7065 line starting around day 90 suggests the presence of a strong ionization source. Also of note is the strengthening of the \ion{Ca}{2} IR triplet, which has become almost as strong as H$\alpha$ by day 165.

\begin{figure*}
\includegraphics[width=7.1in]{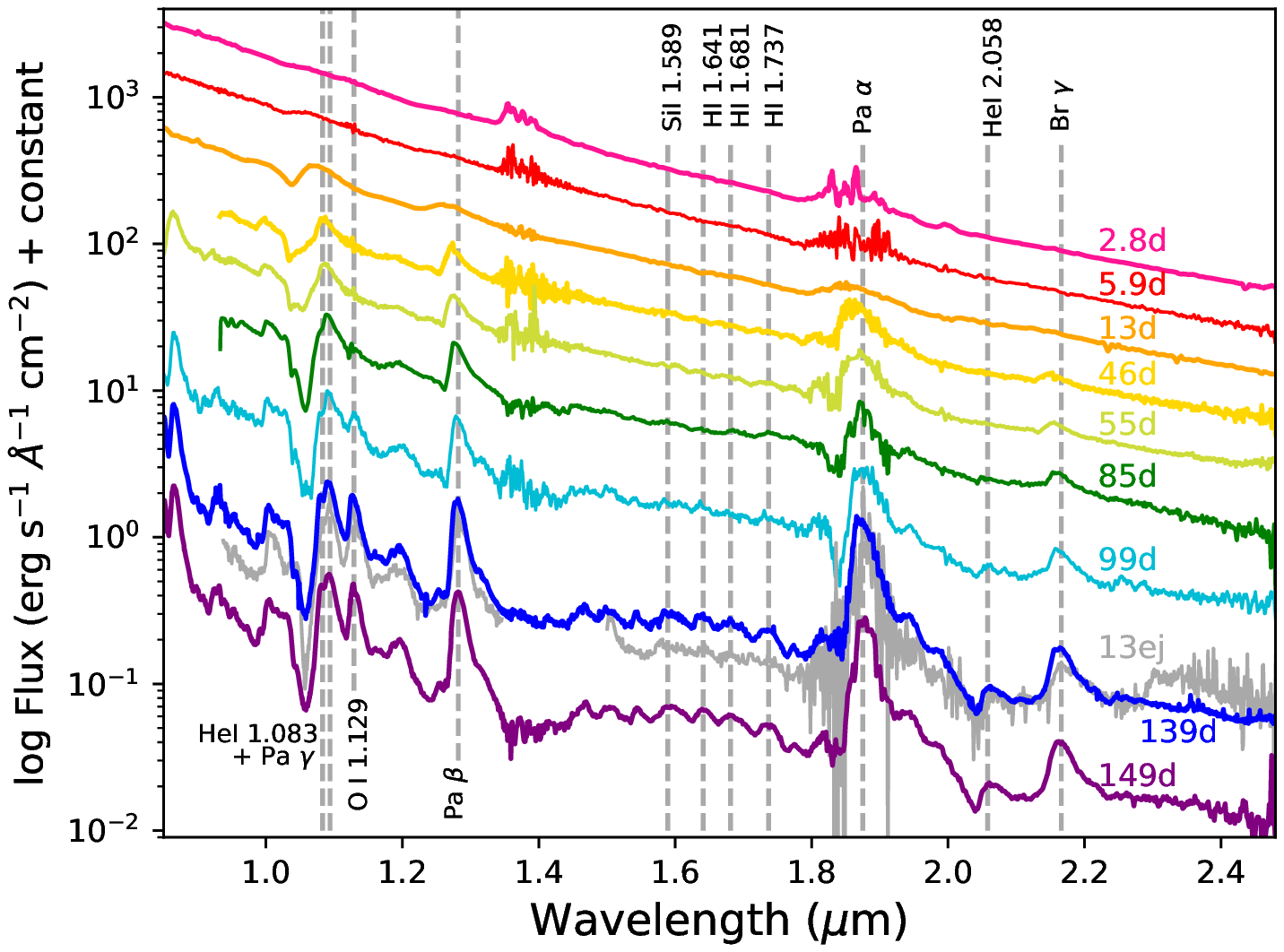}
\caption{NIR spectral sequence of SN~2017gmr. The strongest lines have been marked.  Also shown is the day 138 spectrum of SN2013ej \citep{yuan13ej} which displays prominent CO overtone bands between 2.3--2.5 $\mu$m. A NIR spectroscopy log is presented in Table~\ref{tab:nirspec}.}
\label{fig:IRspec}
\end{figure*}

\subsection{IR Spectra}

Figure \ref{fig:IRspec} shows the NIR spectral evolution from 2--149 days.  Overall the spectra show a decrease in flux with increase in wavelength, typical of young CCSNe.  The spectra from the first week are featureless (minus atmospheric absorption), but by day 13 some Pa $\alpha$ emission begins to emerge. Over the next month Pa $\beta$, and Br $\gamma$ appear as the continuum flux decreases. Both  \ion{He}{1} 1.083 and Pa $\gamma$ are present, although slightly blended. As the SN drops from the plateau phase after 100 days, additional lines of \ion{O}{1}, \ion{Si}{1}, \ion{He}{1}, and other weak hydrogen series are seen.  The CO overtone between 2.3--2.5 $\mu$m is not present in our last two spectra as has been seen for other Type II SNe \citep{yuan13ej,Rho17eaw,Sarangi18, 2019ApJ...873..127T}. This may help rule out dust formation, at least in the first 150 days.

\begin{figure}
\includegraphics[width=3.5in]{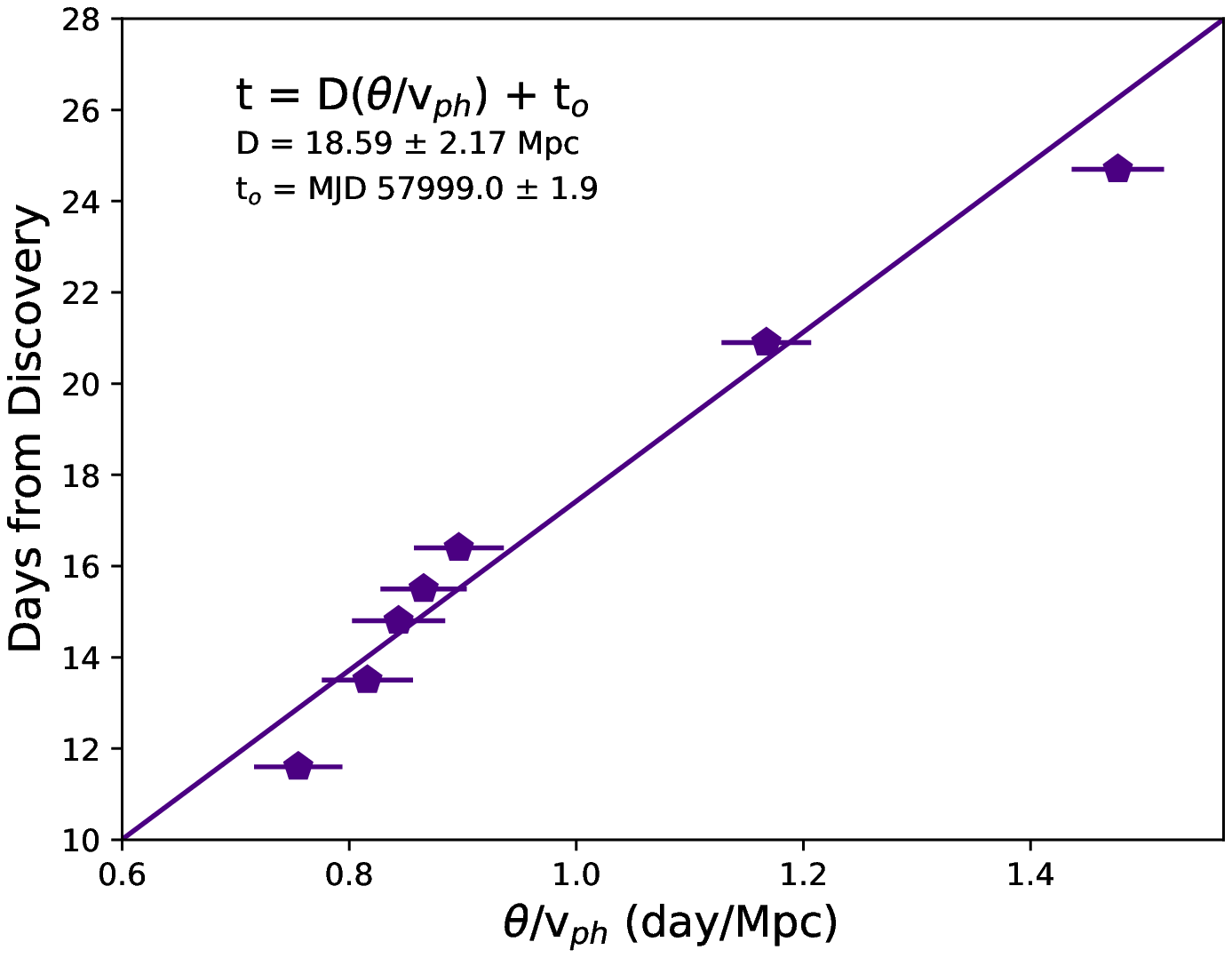}
\caption{Distance determination using EPM for SN~2017gmr.}
\label{fig:EPM}
\end{figure}

\subsection {Distance Measurements}

To help constrain the distance to SN~2017gmr we have used the Expanding Photosphere Method \citep[EPM]{1974ApJ...193...27K}, which relies on the relation between the photometric angular radius and the spectroscopic physical radius of the homologously expanding SN ejecta. Assuming that the outflow is radiating as a diluted blackbody, the observed SN magnitudes are fitted to a blackbody function multiplied by dilution factors, to derive the color temperature and the angular radius. Dilution factors based on atmosphere modeling of Type II SNe were adopted from \citet{2005A&A...439..671D}. Further, to eliminate the effect of filter response function ingrained in the observed broadband magnitudes, the response function is convolved with the blackbody model flux. The convolved function can be expressed in terms of the color temperature and the coefficient values taken from \citet{2001ApJ...558..615H}. Following the same procedure undertaken in \citet{SN2015ba},  expansion velocities were calculated using the \ion{He}{1} $\lambda$5876 and \ion{Fe}{2} $\lambda$5169 lines over the first 50 days of evolution.

The distance is derived from a linear fit to the data in the form of:
\begin{equation}
    t = D(\theta/v_{ph}) + t_{o},
\end{equation}
\noindent
where the slope is the distance, and the y-intercept the date of explosion.  This fit is shown in Figure \ref{fig:EPM}.  From this method we obtain an EPM distance of 18.6 $\pm$ 2.2 Mpc, a value consistent with the 19.1 Mpc used throughout the paper.  It also indicates an explosion epoch of  MJD 57999.0 $\pm$ 1.9 days, which agrees well with the constrained explosion date discussed above.

We have also measured the distance using the Standard Candle Method (SCM), which was first proposed by \citet{2002ApJ...566L..63H} and later expanded on by other authors. SCM uses photometric magnitudes and expansion velocities at 50 days.  For SN~2017gmr these values are: $m_V$ = 14.57 $\pm$ 0.04, $m_R$ = 13.86 $\pm$ 0.02, $m_I$=13.56 $\pm$ 0.03, and $v_{FeII}$ = 5600 km s$^{-1}$.  From these values we get SCM distances (in Mpc) of 16.10 \citep{2005coex.conf..535H}, 16.88 \citep{2006MNRAS.372.1735T}, 24.70 \citep{2006ApJ...645..841N}, 14.38 \citep{2009ApJ...694.1067P}, 10.24 \citep{2017ApJ...835..166D}, and 13.31 \citep{2018A&A...611A..25G}.  Except for \citet{2006ApJ...645..841N}, all other SCM distances are systematically lower than the EPM and kinematic distances.  The same was found for SN~2017eaw and SN~2004et in \citet{Szalai19}, and could be due to CSM-interaction or asymmetries.  The SCM method relies on a correlation between the magnitude and expansion velocity at day 50, which could break down under these conditions. 

\section{Discussion} \label{discussion}

\subsection{$^{56}$Ni Mass} \label{Nickel}

To estimate the $^{56}$Ni mass we employ various methods from the literature, in particular those of \citet{Hamuy2003}, \citet{Jerkstrand2012}, and \citet{2015ApJ...799..215P}. These methods all rely on bolometric luminosities in the radioactive tail phase, so we use the constructed bolometric lightcurve discussed above (Figure \ref{fig:bolo}).  This results in measured $^{56}$Ni masses of 0.130 $\pm$ 0.026 M$_{\sun}$, 0.124  $\pm$ 0.026 M$_{\sun}$, and 0.090 $\pm$ 0.030 M$_{\sun}$ respectively for the three techniques.  In the \citet{2015ApJ...799..215P} calculation, we extrapolated the bolometric luminosity to day 200, and obtain an $L_{bol}$ = 1.85 $\pm$ 0.9 $\times$ 10$^{41}$ erg s$^{-1}$.

Other than SN~1992H, for which the actual $^{56}$Ni mass could be as low as 0.06 M$_{\sun}$ depending on the distance used, and SN~1992am \citep{2003ApJ...582..905H},  this is one of the highest $^{56}$Ni masses reported for normal Type II SNe \citep{2019arXiv190600761A}, higher if there is incomplete gamma photon trapping or if the SN is at a further distance than 19.6 Mpc, lower if there is CSM interaction or if the SN is closer. According to \citet{2017ApJ...841..127M}, less than 5$\%$ of Type II-P SNe have $^{56}$Ni masses as large as 0.12 M$_{\sun}$. For comparison, other ``normal" Type II-P SNe such as  SNe 1999em, 2003gd, and 2004dj each have $^{56}$Ni masses $\sim$0.02 M$_{\sun}$, or a full order of magnitude lower than estimated here \citep{2003MNRAS.338..939E,2005MNRAS.359..906H,2006MNRAS.369.1780V}.  
  
We can also estimate the $^{56}$Ni mass using a steepness factor $S$, where $S = -dM/dt$,  a measure of the transition between the plateau and radioactive tail phases \citep{2003A&A...404.1077E}. Generally an anticorrelation exists, where the steeper the transition, the lower the $^{56}$Ni mass.  Following Equation 7 in \citet{2018MNRAS.480.2475S} we measure a steepness factor $S = 0.070 \pm 0.007$ mag d$^{-1}$ , which corresponds to an estimated $^{56}$Ni mass of $\sim$ 0.055  M$_{\sun}$.  This is significantly smaller than the value obtained using the late-time bolometric luminosity, and more consistent with other normal Type II-P SNe.  This inconsistency could be due to the degree of mixed $^{56}$Ni in the ejecta, since the same amount of $^{56}$Ni will create a steeper decline if it is centrally located rather than mixed. The mixed $^{56}$Ni will actually increase the radiative diffusion timescale, causing the transition to appear shallower. 

\begin{figure}
\includegraphics[width=3.5in]{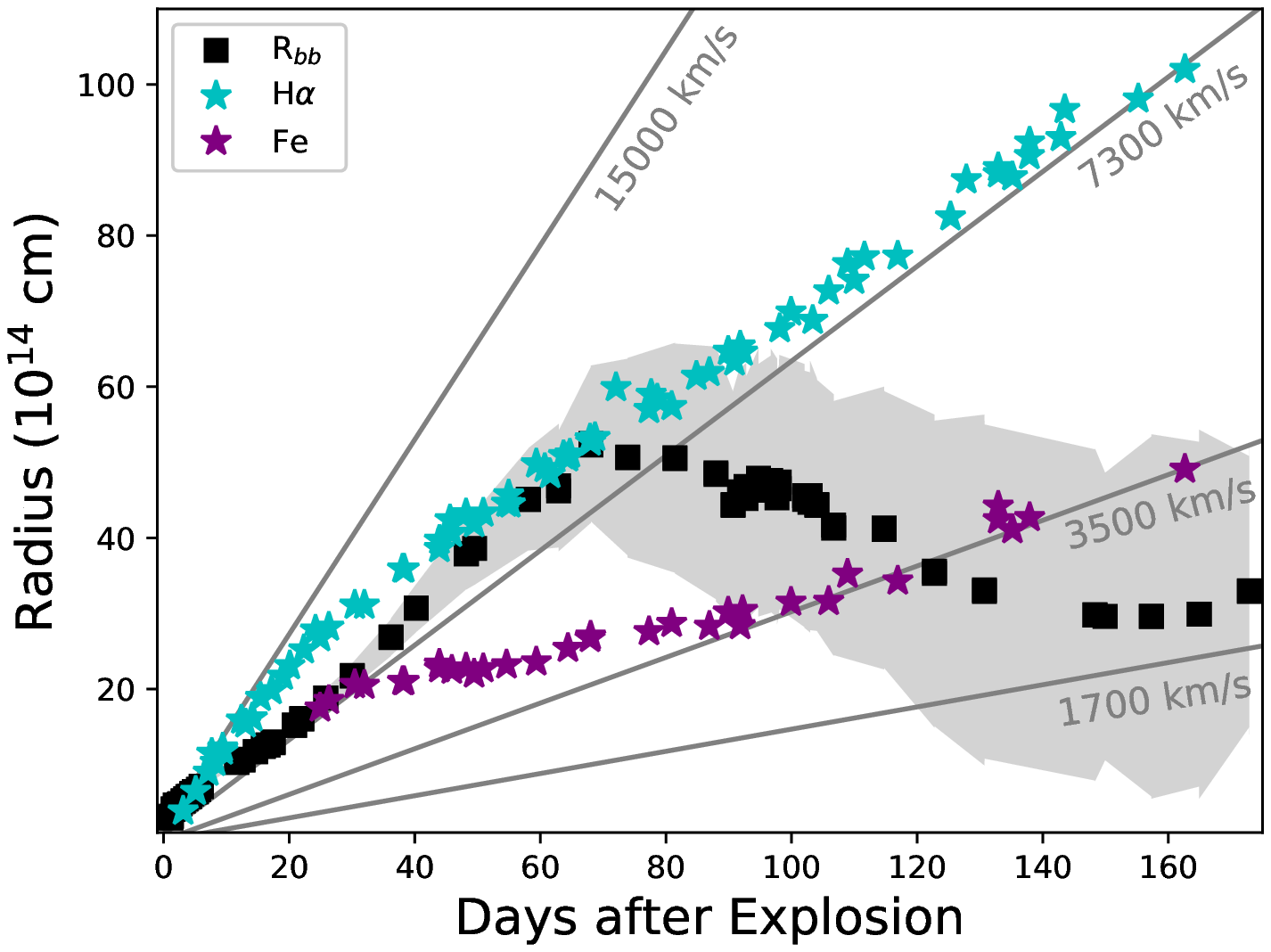}
\caption{Blackbody radius of SN~2017gmr compared to the photospheric radius calculated from H$\alpha$ and \ion{Fe}{2} $\lambda$5169.  Lines of constant velocity are plotted in dark gray, while the uncertainty in R$_{bb}$ is shown in light gray.}
\label{fig:linevelocity}
\end{figure}

\begin{figure}
\includegraphics[width=3.5in]{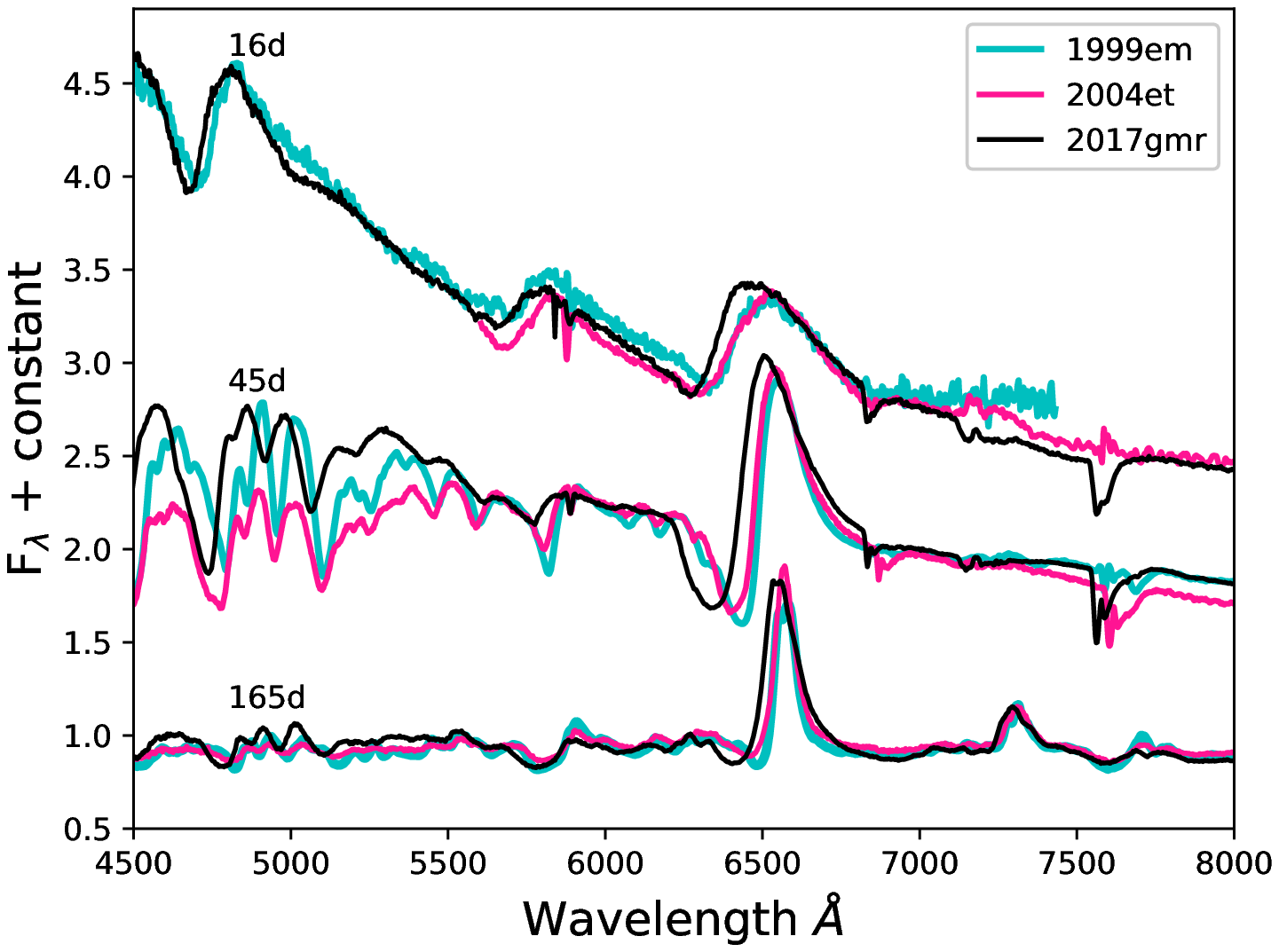}
\caption{Comparison of SN~2017gmr with other well-studied Type II-P SNe 1999em and 2004et at various epochs.  Data from \citet{2002PASP..114...35L}, \citet{Faran2014}, and \citet{2006MNRAS.372.1315S} and obtained from WISeREP \citep{Yaron12}.}
\label{fig:etemcomp}
\end{figure}

\subsection{Extremely fast ejecta}\label{fast}
In Figure \ref{fig:linevelocity} we show the evolution of the line velocities of both H$\alpha$ and \ion{Fe}{2} $\lambda$5169 (shown as a function of radius over time). H$\alpha$ falls from 15000 km s$^{-1}$ near explosion to a relatively stable value of 7000 - 8000 km s$^{-1}$ during the radioactive tail. \ion{Fe}{2} $\lambda$5169, a more reliable measurement of photospheric velocity than H$\alpha$, settles to a late-time velocity of 3500 km s$^{-1}$.  These expansion velocities are higher than average for Type II SNe, and for Type II-P SNe in particular.  In Figure \ref{fig:etemcomp} we show the comparison of SN~2017gmr optical spectra at various epochs with the well-studied SN~1999em and SN~2004et. At all epochs the line velocities of SN~2017gmr are faster than those of the other two.

\begin{figure*}
\includegraphics{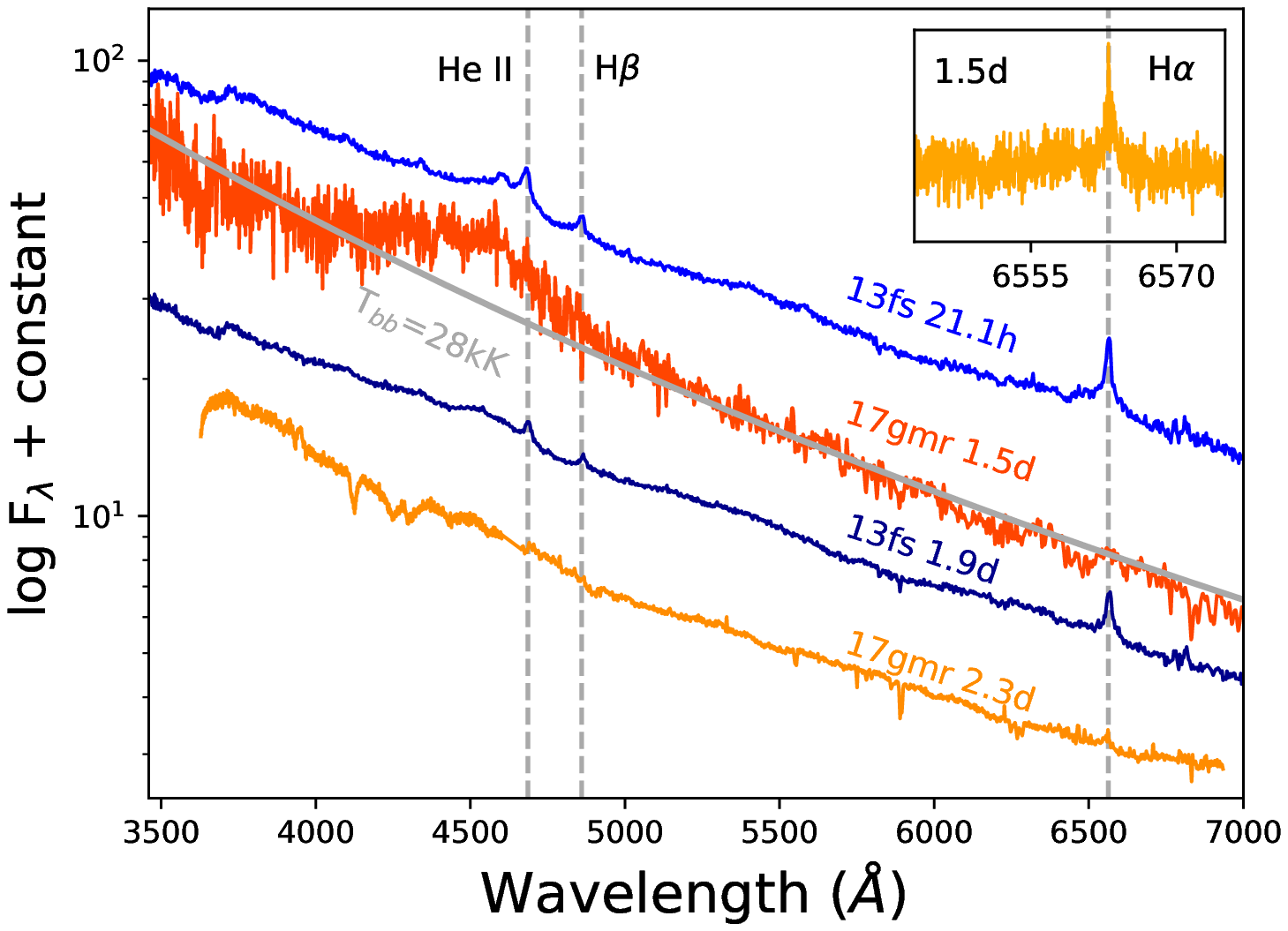}
\caption{Comparison of the earliest spectroscopy of SN~2017gmr with that of SN~2013fs from \citet{Yaron2017}. The spectrum in the inset is the 1.5d Keck HIRES spectrum of SN~2017gmr, which shows a weak H$\alpha$ feature with a width $\sim$55 km s$^{-1}$.  This feature is also weakly seen in the 2.3d HET spectrum; the 1.5d FLOYDS spectrum is too low resolution to identify any narrow H$\alpha$. A 28kK blackbody is also plotted over the FLOYDS spectrum in gray. }
\label{fig:flash}
\end{figure*}

From \citet{Gutierrez2017a}, the mean velocities on day 53 for a sample of 122 Type II SNe ((measured from the absorption minimum) are 6365 km s$^{-1}$ and 3537 km s$^{-1}$ for H$\alpha$ and \ion{Fe}{2} $\lambda$5169, respectively. In comparison, SN~2017gmr has velocities on day 53 of  9330 km s$^{-1}$ and 5240 km s$^{-1}$ for H$\alpha$ and \ion{Fe}{2} 5169.   By day 115, the difference in the \ion{Fe}{2} $\lambda$5169 velocities has decreased, 2451 km s$^{-1}$ average versus 3520 km s$^{-1}$ for SN~2017gmr, but H$\alpha$ remains almost 2000 km s$^{-1}$ faster than the mean value of 5805 km s$^{-1}$.

The faster line velocities seem to correlate well with the high inferred $^{56}$Ni mass and maximum luminosity of SN~2017gmr. \citet{Gutierrez2017b} found a correlation between expansion velocities and $^{56}$Ni mass that indicated that more energetic explosions (resulting in faster expansion velocities) created higher $^{56}$Ni mass. When combined with previous conclusions of \citet{2002ApJ...566L..63H}, \citet{Hamuy2003} and \citet{2015ApJ...799..215P}, this suggests that the more energetic the explosion, the higher the luminosity, expansion velocity, and $^{56}$Ni production. This may suggest that SN~2017gmr had an unusually energetic explosion,  although low ejecta mass can also allow for high ejecta velocities.

There are other ways to create faster line velocities.  If CSM interaction is occurring it can excite H$\alpha$ and other lines at larger radii (and therefore higher velocities).  This means that lines which would have otherwise already recombined in the outer, faster parts of the ejecta will be reionized and give the appearance of faster ejecta at later times.  Faster expansion velocities can also arise from asymmetries in the explosion. \citet{2011MNRAS.415.3497D} found that asphericities in the ejecta  of an axially symmetric explosion can change the location of the P-Cygni minimum with inclination as much as 30$\%$ in the photospheric phase. We explore the possibility of an asymmetric explosion in Section \ref{asymmetry}.

\subsection{Early Narrow Features?} \label{flash}
 
Narrow lines seen within the first few days of explosion can be useful to infer composition, velocity, and density of the CSM surrounding the SN progenitor \citep{GalYam2014}.  One of the most well known objects displaying this phenomenon,  SN~2013fs,  showed  narrow ($\sim$ 100 km s$^{-1}$) lines of oxygen, helium, and nitrogen within the first few hours of explosion \citep{Yaron2017, 2018MNRAS.476.1497B}.  These high excitation lines disappeared over the next two days, and eventually the spectra resembled that of a normal Type II SN. Similar behavior has been seen in SN~1983K \citep{1985ApJ...289...52N}, SN~2006bp \citep{2007ApJ...666.1093Q}, SN~2013cu \citep{GalYam2014}, SN~1998S \citep{Shivvers2015}, PTF11iqb \citep{PTF11iqb}, SN~2016bkv \citep{Hosseinzadeh18}, and SN~2014G \citep{2016MNRAS.462..137T}. \citet{Khazov16} found  14\% to 18\% of their sample of SNe II showed signs of early narrow lines, which they conclude is a lower limit for the SNe II population as a whole.

These narrow lines were interpreted as the flash ionization of a WR-like wind for SN~2013cu \citep{GalYam2014}.  Later interpretation suggested that it instead possibly the ionization of the cool dense wind from an LBV/YHG progenitor \citep{2014A&A...564A..30G} which is more consistent with a type IIb SN progenitor.  Furthermore, \citet{PTF11iqb} found that PTF11iqb had a RSG progenitor and the early narrow lines were likely the result of shock ionization from CSM interaction. A similar conclusion about the progenitor of SN~1998S was also reached in \citet{Shivvers2015} and \citet{Mauerhan12}. In other words, WR-like wind features (particularly of hydrogen rich WNH type) can be seen in early spectra if there are enough high energy photons to fully ionize the progenitor's cool dense wind.

SN~2017gmr was observed spectroscopically within hours after discovery, and likely within 1.5 days of shock breakout, yet the only narrow emission line seen was that of H$\alpha$ (Figure \ref{fig:flash}), and only with the higher resolution instruments.  The Keck HIRES spectrum on day 1.5 (inset of Figure \ref{fig:flash}) shows a narrow H$\alpha$ emission with a Gaussian FWHM velocity of $\sim$55 km s$^{-1}$. This is suggestive of a RSG wind (see \citealt{SmithMassLoss14}). The spectral resolution of this data is $\sim$ 7 km s$^{-1}$, so the velocity of the ionized material is fully resolved.  For reference, SN~1998S was observed with the same instrument 1.86 days after discovery and had a narrow component velocity of $\sim$40 km s$^{-1}$ \citep[][albeit with lines other than H$\alpha$ also present]{Shivvers2015}. The day 2.3 HET spectrum also seems to show a narrow but weak H$\alpha$ feature with a moderately higher intermediate-width FWHM velocity of $\sim$1000 km s$^{-1}$. The broadening of the line may be due to electron scattering in the CSM, and the narrow feature may be embedded within, but it has likely faded by this epoch. This feature is completely gone in the HET spectrum 3 days later; in its place is a broad, blueshifted H$\alpha$ emission with an expansion velocity of 15000 km s$^{-1}$.

One other noticeable feature in the very early spectra is the broad emission around $\sim$ 4600 \AA\  (see Figure~\ref{fig:flash}).  A similar broad bump was seen in SN~2006bp \citep{2007ApJ...666.1093Q} and SN~2013fs \citep{2018MNRAS.476.1497B} and was attributed to blueshifted \ion{He}{2} $\lambda$4686 formed from the SN ejecta beneath a CSM shell. These two objects did also show narrow \ion{He}{2} $\lambda$4686 emission on the red edge of the broad 15000 km s$^{-1}$ line, which is absent in SN~2017gmr.

The lack of narrow high-ionization lines in the early spectra would seem to suggest that if nearby CSM was present, its density was too low to yield detectable emission. Alternatively, it could imply that the photons were not energetic enough to doubly ionize He in the CSM, even if SN~2017gmr likely had a very energetic explosion. If the CSM density was adequately high, this too could prevent narrow lines from forming, as it would self-absorb all of the high energy photons.  Another option would be that the narrow, high-ionization lines were present before our first spectrum at 1.5 days, but were produced from asymmetric CSM which was quickly enveloped by the spherically expanding SN ejecta \citep{PTF11iqb}.  We will discuss this possibility further in the next section. 

Another luminous Type II, SN~2016esw, was also caught within a day of explosion and  showed no signs of high-ionization emission lines \citep{SN2016esw}. The authors conclude that the progenitor of SN~2016esw was likely surrounded by low-density CSM some distance away from the surface of the star that eventually showed signs of interaction 2-3 weeks after explosion. Similarly, the type II-P SN~2017eaw did not show early flash signatures \citep{2019ApJ...875..136V}, except for possibly the $\sim$160 km s$^{-1}$ H$\alpha$ line seen by \citet{2019MNRAS.485.1990R} on day 2.5.  Unlike SN~2016esw though, neither SN~2017eaw nor SN~2017gmr showed obvious signs of CSM interaction in the shape of the H$\alpha$ emission line the first few weeks after explosion.
 
\subsection{Circumstellar Interaction or Asymmetric Explosion?} \label{asymmetry}

 \begin{figure}
\includegraphics[width=3.6in]{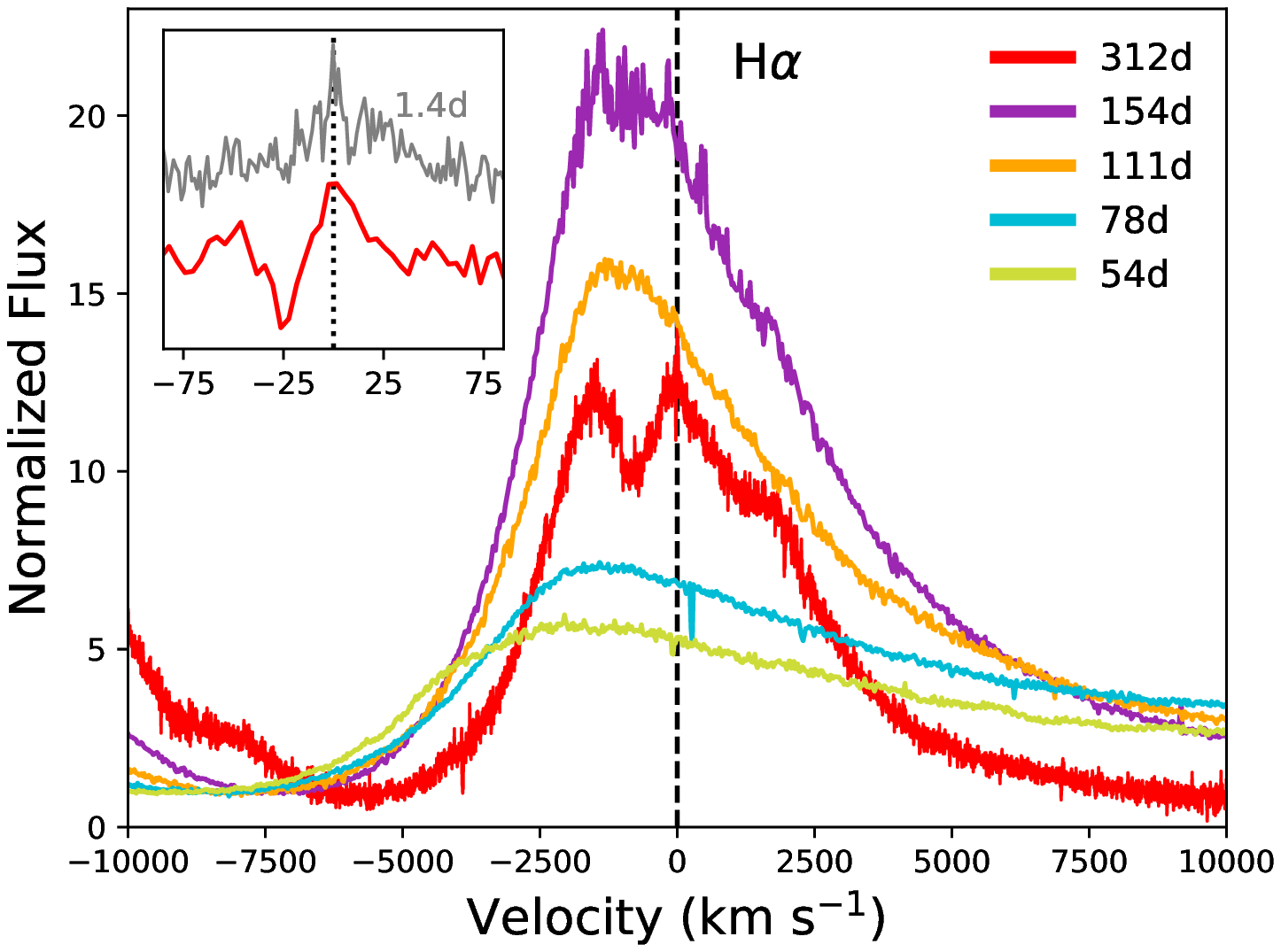}
\includegraphics[width=3.6in]{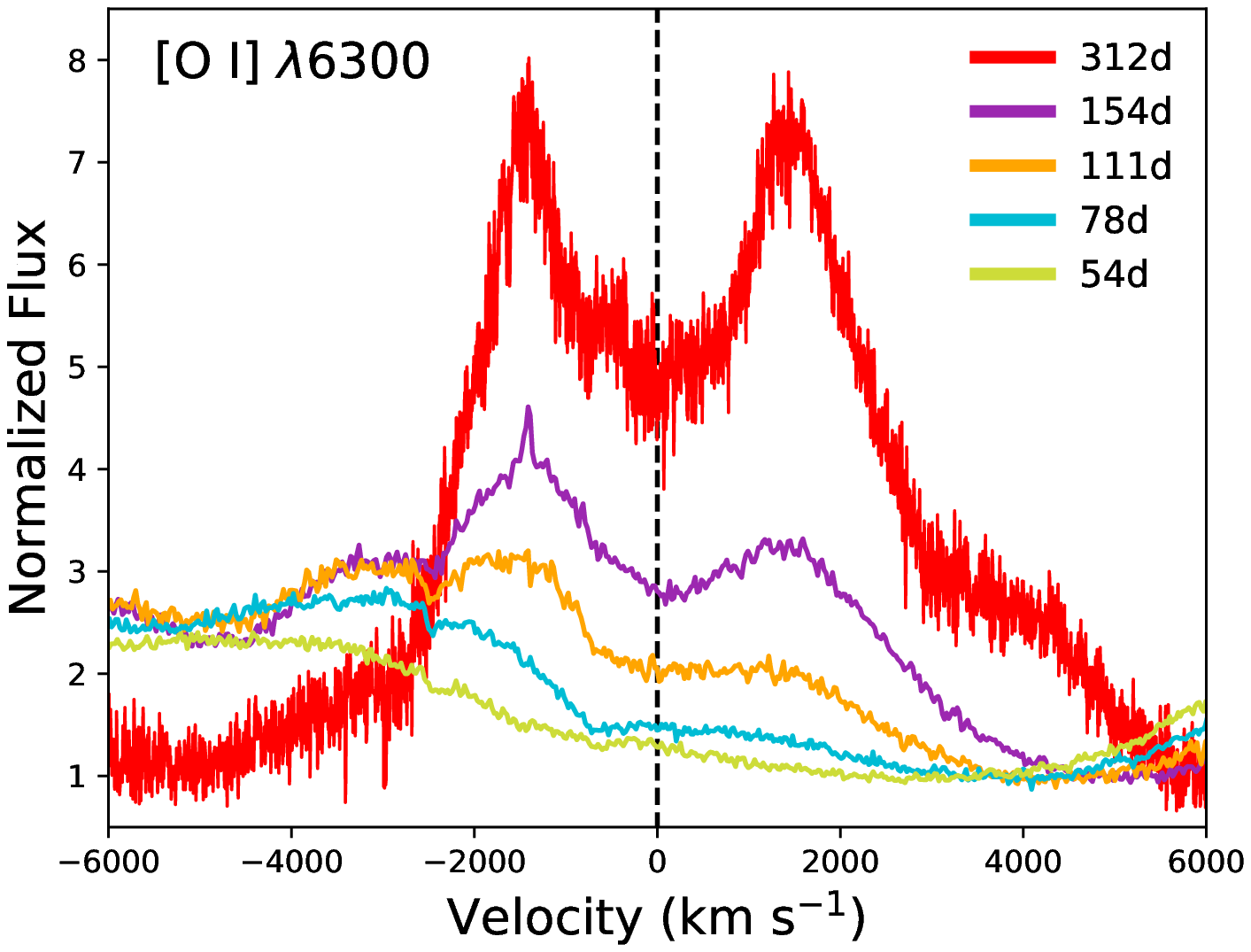}
\caption{Evolution of the H$\alpha$ (top) and [OI] $\lambda\lambda$6300,6363 \AA\ (bottom) emission lines from our moderate and high-resolution spectra.  The lines have been normalized to the minimum of the H$\alpha$ P-Cygni line.  The multi-peaked shape begins to arise between 110-150 days, but is clearly evident by our last spectrum on day 312.}
\label{fig:Haevo}
\end{figure}

When the SN reappeared from behind the Sun in 2018 July we obtained one high-resolution echelle spectrum with MIKE on Magellan/Clay on day 312.  The late-time analysis on SN~2017gmr is beyond the scope of this paper, and will be discussed in depth in an upcoming paper, but due to the implications for the early time evolution we are including the H$\alpha$ and [\ion{O}{1}] lines here.  In Figure \ref{fig:Haevo} we show in red the day 312 spectrum compared to the other moderate-resolution MMT spectra.  Instead of a single broad line, H$\alpha$ clearly shows three intermediate width peaks. The same is seen in the [\ion{O}{1}] doublet, albeit the right peak of the  6300 \AA\ line is stronger than the red peak of H$\alpha$ due to the overlap of the blue peak from the  6364 \AA\ line of the doublet. Signs of this asymmetry can even be seen in the day 154 spectrum.

\begin{figure}
\includegraphics[width=3.6in]{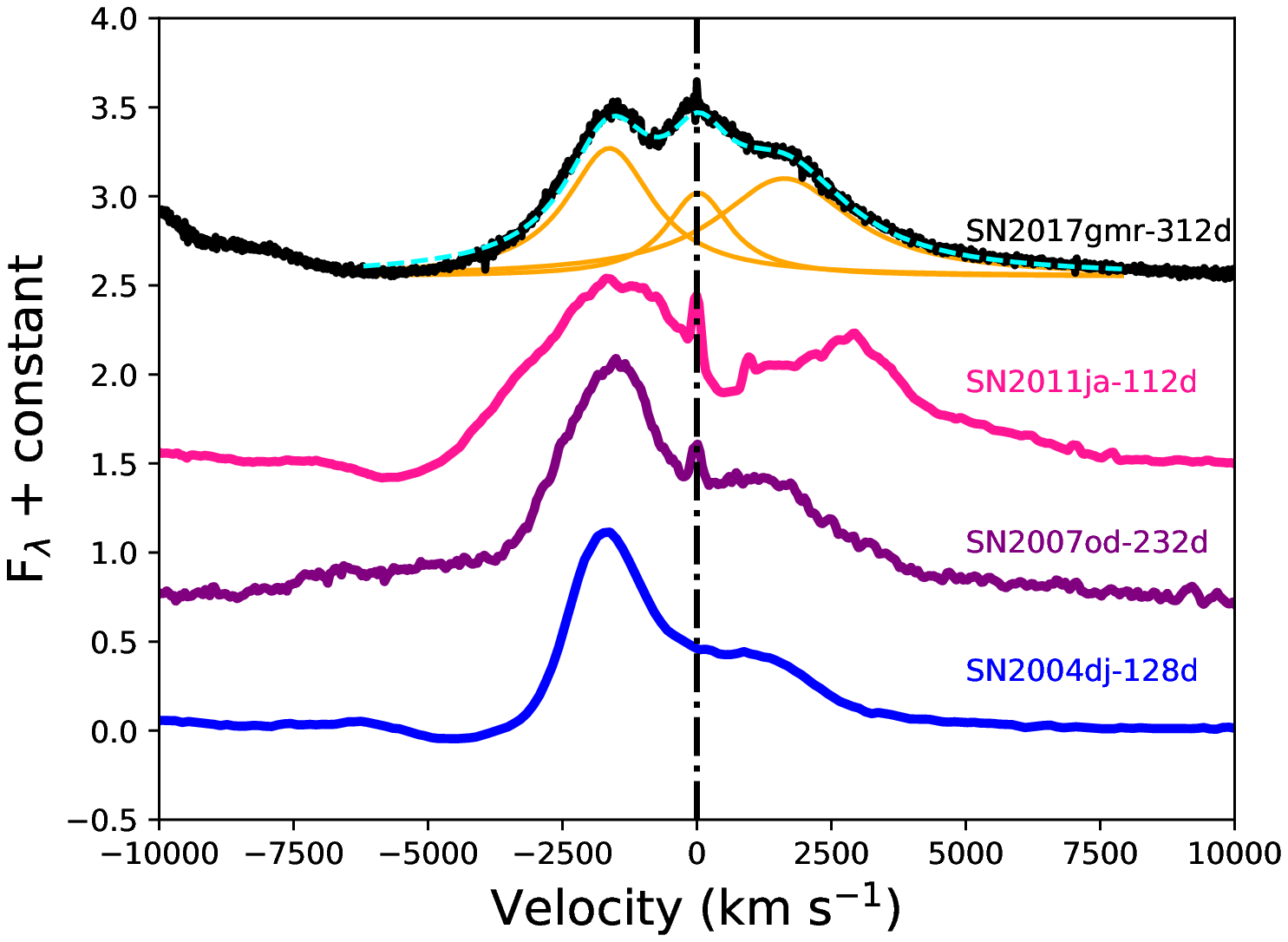}
\caption{SN~2017gmr compared with other Type II-P SNe showing signs of CSM interaction during the radioctive tail phase.  Data are from \citet[SN2011ja]{SN2011ja}, \citet[SN2007od]{Andrews10}, and \citet[2004dj]{2006MNRAS.369.1780V}. Lorentzian fits to the multiple components of SN~2017gmr are shown in orange, and the total fit of the seperate components are overplotted as a dashed cyan line.}
\label{fig:csmha}
\end{figure}

In Figure \ref{fig:csmha} we show that the multi-peaked H$\alpha$ can be fit with three Lorentzians, one centered at 0 km s$^{-1}$ (6563 \AA) and blue and red peaks at roughly $\pm$ 1700 km s$^{-1}$ ($\pm$ 35 \AA).  The same velocities are seen in [\ion{O}{1}], for both the $\lambda$6300 and $\lambda$6363 lines, but the doublet nature of the line makes it appear distinctly different. The red peak of $\lambda$6300 would fall at $\sim$ 6335 \AA\ while the blue peak of $\lambda$6364 would fall around $\sim$ 6330 \AA\,  making the red peak of $\lambda$6300 seem as bright as the blue peak, and swamping the emission at the center of the line.

First presented in \citet{2005Sci...308.1284M}, double-peaked emission lines in SN spectra are often interpreted as ejecta interacting with asymmetric CSM, most commonly in a disc or torus \citep{Hoffman08,2008Sci...319.1220M,2009MNRAS.397..677T,Mauerhan14,PTF11iqb,Andrews17}. In this scenario, the underlying  broad component traces emission from the free expansion of the SN ejecta, while the intermediate components are formed in the post-shock region between the forward and reverse shocks created as the ejecta crashes into the CSM. When the fast moving SN ejecta collides with the slow moving CSM, depending on the density of surrounding material, the CSM can be accelerated from speeds of 10 - 100 km s$^{-1}$ up to thousands of km s$^{-1}$. The red and blue peaks therefore are the result of the ejecta accelerating the CSM material radially outward from the explosion. In the case of SN~2017gmr the CSM was likely accelerated from a normal RSG wind speed of $\sim$ 55 km s$^{-1}$ to the observed intermediate feature speed of $\sim$ 1700 km s$^{-1}$. Examples of other Type II SNe at somewhat similar phases as SN~2017gmr showing multi-peaked H$\alpha$ are shown in Figure \ref{fig:csmha}.  

The fact that we do not see narrow emission lines does not necessarily discount the possibility of SN~2017gmr being a partially CSM-interaction powered event. CSM interaction can be inferred based on the intermediate-width line shapes and velocities. As explained in \citet{PTF11iqb}, \citet{SNHandSmith}, and \citet{iptf14hls}, a disc-like geometry in the CSM may allow the CSM interaction to be hidden below the photosphere after the disc is enveloped by the fast SN ejecta.  If the region of CSM interaction is happening below the ejecta photosphere, and the CSM is sufficiently dense, it can be hidden for long periods of time because the sustained CSM interaction luminosity itself keeps the surrounding SN ejecta ionized and optically thick. This could help explain the extended high-luminosity of SN~2017gmr.  All that is required is that the disc or torus of material has a limited radial extent (i.e. $\leq$ 100 AU) so that it can be overrun early by the SN photosphere. Only when the photosphere recedes internal to the CSM location (which has been pushed outward to 1700 km s$^{-1}$ due to the Doppler acceleration)  will the intermediate-width lines be revealed. 

If we assume that high-ionization lines were observable prior to our 1.5 day spectrum we can use the expansion velocity of H$\alpha$ (15000 km s$^{-1}$) to infer that the outer edge of the CSM must be closer than 1.8 $\times$ 10$^{14}$ cm (or $\sim$ 2500 R$_{\sun}$).  This is roughly the same radius infered for SN~2013cu \citep{GalYam2014} and PTF11iqb \citep{PTF11iqb}.

The other possibility is that the multiple peaks seen in the hydrogen lines could come from asymmetries in $^{56}$Ni in the ejecta. This was the scenario presented for SN~2004dj \citep[shown in Figure \ref{fig:csmha}]{Chugai2005}, SN~2010jp \citep{2012MNRAS.420.1135S} and SN~2016X \citep{2018MNRAS.475.3959H,2019ApJ...873L...3B}. In a forthcoming paper  Nagao et al. (2019) find there is strong polarization in SN~2017gmr indicative of an aspherical explosion.  Non-uniformity of $^{56}$Ni could cause  uneven ionization and excitation in the ejecta, and produce multi-peaked emission lines. SN~2004dj showed strong H$\alpha$ asymmetry immediately after the plateau phase ended, during the epoch of increased polarization  \citep{2006Natur.440..505L}. As we show in Figure \ref{fig:Haevo}, distinct multiple peaks are not present until sometime between 154-312 days, or a significant time period after the end of the plateau.   Also of note is that there is  a component at rest velocity at late times in SN~2017gmr which would have to come from some spherically distributed radioactive material.

In general it is difficult to disentangle the two mechanisms.  The low polarization at early times is explained by Nagao et al. (2019) by the hydrogen envelope hiding a highly asymmetric helium core which is only observable when the optical depth decreases. We suggest it could also be explained partially (or in full) by the spherical symmetry of the hydrogen envelope erasing the polarization signatures of deeply embedded asymmetric CSM interaction.  The deviation from $^{56}$Co decay in the late-time lightcurve can be due to incomplete $\gamma$ photon trapping caused by a non-spherical ejecta, or it could be due to a decrease in the shock interaction. Whatever the mechanism, the emission line shapes emerging during the nebular phase indicate a deviation from spherical symmetry, whether it be from asymmetric stellar ejecta  or shock interaction with a disc or torus of CSM.

\subsection{Dust Formation?}

As we briefly mention above, the lightcurves shown in Figure \ref{fig:fulllc} show that there is a clear discrepancy between the observed late-time luminosity and what is expected due to $^{56}$Co decay.  The fast decline could indicate the halting of shock interaction as a primary energy source, or that there is incomplete trapping of gamma rays as we discuss in Section \ref{Nickel}.  It could also be due in all, or part, to dust formation in the ejecta.

Along with a decrease in optical luminosity from the growth of dust grains, we can also expect to see a blue shifted asymmetry in the optical emission lines since dust in the ejecta would attenuate the receding red side of the SN more than the blue.  First detected in SN~1987A \citep{1989LNP...350..164L}, evidence for dust formation has been seen in many CCSNe including SN~2003gd \citep{Sugerman06}, SN~2004et \citep{2009ApJ...704..306K}, SN~2005ip \citep{2009ApJ...695.1334S,2010ApJ...725.1768F,2012ApJ...756..173S,2019MNRAS.485.5192B}, SN~2006jd \citep{2012ApJ...756..173S}, SN~2007od \citep{Andrews10,2011MNRAS.417..261I}, SN~2010jl \citep{2012AJ....143...17S, 2014Natur.511..326G} and one of the clearest cases SN~2006jc \citep{Smith082006jc,2008MNRAS.389..141M}. In conjunction with the emission line asymmetry and a decrease in the optical light curve, a corresponding increase in the IR luminosity is often observed as new dust grains form in the ejecta. 

It is unlikely that dust has formed in SN~2017gmr by $\sim$ 150d for a few reasons. First, a blackbody fit to the optical and NIR spectroscopy and photometry around day 150 indicates a T$_{bb}$ = 6800 K, a temperature much too high for grain condensation.  Secondly, the bolometric lightcurve also shows a deviation from expected $^{56}$Co decay.  If dust formation was occurring the lightcurves in individual bands will change, but the total bolometric lightcurve would be unchanged.  Finally, as we mention above, the NIR spectroscopy during the early nebular phase fail to reveal the first overtone of CO (Fig. \ref{fig:IRspec}). Normally the detection of CO heralds the formation of dust \citep{Gerardy2000, Sarangi13}.  Therefore the blue-peaked hydrogen emission profiles and the fast decline in L$_{bol}$  is likely due to other physical characteristics of the SN such as asymmetries and CSM interaction, not dust formation.  This does not discount the possibility that in later epochs we may begin to see signatures of dust condensation in the ejecta.

\section{Conclusions} \label{conclusion}
SN~2017gmr is one of the more luminous Type II-P SNe discovered to date, with one of the largest measured $^{56}$Ni masses for a II-P event.  Not only does it peak at M$_{V}$ = $-$18.3 mag, but by 150 days after explosion it has declined less than 3 mag in the $V$-band.  If the late-time luminosity is powered solely by radioactive decay, then the mass of $^{56}$Ni is 0.130 $\pm$ 0.026 M$_{\sun}$, quite massive for a Type II-P SN. The line velocities are abnormally fast for a Type II-P event, which could be due to an extremely energetic explosion,  asymmetries in the ejecta, or CSM interaction reionizing the faster, outer parts of the ejecta.  The inferred progenitor radius is $\sim$500 R$_{\sun}$, on the lower end for a RSG, but within normally expected values.

CSM interaction is an efficient way to convert SN ejecta kinetic energy into radiative luminosity.  The high luminosity of SN~2017gmr at late times and the bump in the early-time $U$ and $B$ lightcurves could both be the result of an added energy contribution from CSM interaction.  The fact that no narrow lines are seen at early times could be due to the spherical ejecta quickly overtaking the asymmetric CSM, and the lack of narrow lines at late times only indicates that the SN shock has moved completely through the close-in CSM. In other words, all the slow moving CSM has been swept up by the shock. Low polarization during the plateau phase (Nagao et al. 2019) could also be explained by mostly spherical ejecta  enveloping a dense, close-in asymmetric CSM. Since these CSM interaction photons are thermalized deep inside the opaque SN ejecta envelope, their polarization signature from asymmetric CSM would be erased.  Asymmetric explosions producing jets or blobs of $^{56}$Ni could also create the asymmetric emission lines and the high line velocities.

SN~2017gmr was caught very young, and the collection of high-cadence multiwavelength data began immediately.  This has allowed us the ability to not only explore the early behavior of Type II SNe, but the years of mass loss prior to explosion.  More instances of early data are needed to understand both this mass loss, and the diversity among SNe in these early time properties. Either SN~2017gmr is an unusually energetic Type II-P SN explosions, or it has the assistance of CSM interaction and asymmetries to make it appear so.  Continued observations of SN~2017gmr are ongoing and are necessary to help disentangle the various energy inputs and the overall geometry of this unique event.

\acknowledgements

We acknowledge contributions from S. Taubenberger,  T. Nagao, and N. Suntzeff.  JEA and NS receives support from NSF grant AST-1515559. Research by DJS and SW is supported by NSF grants AST-1821987 and 1821967. 
Research by SV is supported by NSF grants AST-1813176. GH, DAH, and CM were supported by NSF grant AST-1313484.  KM acknowledges the support from Department of Science and Technology (DST), Govt. of India and Indo-US Science and Technology Forum (IUSSTF) for the WISTEMM fellowship and Dept. of Physics, UC Davis where a part of this work was carried out. G.H thanks the LSSTC Data Science Fellowship Program, which is funded by LSSTC, NSF Cybertraining Grant \#1829740, the Brinson Foundation, and the Moore Foundation; his participation in the program has benefited this work. J.V. and his group at Konkoly are supported by the project “Transient Astrophysical Objects” GINOP-2-3-2-15-2016-00033 of the National Research, Development and Innovation Office (NKFIH), Hungary, funded by the European Union. Lingzhi Wang is sponsored (in part) by the Chinese Academy of Sciences (CAS), through a grant to the CAS South America Center for Astronomy (CASSACA) in Santiago, Chile. MB acknowledges support from the Swedish Research Council (Vetenskapsr\aa det), the Swedish National Space Board and the research environment grant ``Gravitational Radiation and Electromagnetic Astrophysical Transients (GREAT)''. JS acknowledges support from the Packard Foundation.  O.R. acknowledge support by projects IC120009 ``Millennium Institute of Astrophysics (MAS)'' of the Iniciativa Cient\'ifica Milenio del Ministerio Econom\'ia, Fomento y Turismo de Chile and CONICYT PAI/INDUSTRIA 79090016. Brajesh Kumar also acknowledges the Science and Engineering Research Board (SERB) under the Department of Science \& Technology, Govt. of India, for financial assistance in the form of National Post-Doctoral Fellowship (Ref. no. PDF/2016/001563). Research by JCW is supported by NSF AST-1813825. Support for F.O.E is provided by the Ministry of Economy, Development, and Tourism's Millennium Science Inititative through grant IC120009, awarded to the Millennium Institute of Astrophysics, MAS. KM acknowledges support from H2020 through an ERC Starting Grant (758638). F.O.E acknowledges support from the FONDECYT grant 11170953. MG is supported by the Polish NCN MAESTRO grant 2014/14/A/ST9/00121. SB is partially supported by PRIN INAF 2017 “Towards the SKA and CTA era: discovery, localisation and physics of transient sources (PI M. Giroletti).” M. D. S. is funded by a project grant (8021-00170B) from the Independent Research Fund Denmark and also in part by a generous grant (13261) from VILLUM FONDEN. N.E.R. acknowledges support from the Spanish MICINN grant ESP2017-82674-R and FEDER funds. E.Y.H., C.A., S.D. and M.S. acknowledge the support provided by the National Science Foundation under grant No. AST-1613472. CG was supported by a grant from VILLUM FONDEN (project number 16599). CG was supported by a grant from VILLUM FONDEN (project number 16599). X.W. is supported by the National Natural Science Foundation of China (NSFC grants 11325313, 11633002, and 11761141001), and the National Program on Key Research and Development Project (grant no. 2016YFA0400803).  This work was also partially Supported by the Open Project Program of the Key Laboratory of Optical Astronomy, National Astronomical Observatories, Chinese Academy of Sciences. NUTS is supported in part by IDA (Instrumentation center for Danish Astrophysics)

Observations reported here were obtained at the MMT Observatory, a joint facility of the University of Arizona and the Smithsonian Institution. This work makes use of data obtained by the Las Cumbres Observatory network. Based on observations obtained at the Gemini Observatory  under program GN-2017B-Q-52 (PI: Sand), which is operated by the Association of Universities for Research in Astronomy, Inc., under a cooperative agreement with the NSF on behalf of the Gemini partnership: the National Science Foundation (United States), National Research Council (Canada), CONICYT (Chile), Ministerio de Ciencia, Tecnolog\'{i}a e Innovaci\'{o}n Productiva (Argentina), Minist\'{e}rio da Ci\^{e}ncia, Tecnologia e Inova\c{c}\~{a}o (Brazil), and Korea Astronomy and Space Science Institute (Republic of Korea). Based in part on observations collected at Copernico and Schmidt telescopes (Asiago, Italy) of the INAF - Osservatorio Astronomico di Padova. This paper is partially based on observations collected with the 1.22m \textit{Galileo} telescope of the Asiago Astrophysical Observatory, operated by the Department of Physics and Astronomy "G. Galilei" of the University of Padova. Based on observations made with the Nordic Optical Telescope, operated by the Nordic Optical Telescope Scientific Association at the Observatorio del Roque de los Muchachos, La Palma, Spain, of the Instituto de Astrofisica de Canarias. Based on observations obtained at the Southern Astrophysical Research (SOAR) telescope, which is a joint project of the Minist\'{e}rio da Ci\^{e}ncia, Tecnologia, Inova\c{c}\~{o}es e Comunica\c{c}\~{o}es (MCTIC) do Brasil, the U.S. National Optical Astronomy Observatory (NOAO), the University of North Carolina at Chapel Hill (UNC), and Michigan State University (MSU). Partially based on observations collected at the European Southern Observatory under ESO programme 099.D-0543(A). We thank the staff of IAO, Hanle and CREST, Hosakote, that made the observations from HCT possible. The facilities at IAO and CREST are operated by the Indian Institute of Astrophysics, Bangalore.

The authors wish to recognize and acknowledge the very significant cultural role and reverence that the summit of Mauna Kea has always had within the indigenous Hawaiian community.  We are most fortunate to have the opportunity to conduct observations from this mountain. Some of the data presented herein were obtained at the W.M. Keck Observatory, which is operated as a scientific partnership among the California Institute of Technology, the University of California and the National Aeronautics and Space Administration. The Observatory was made possible by the generous financial support of the W.M. Keck Foundation. Visiting Astronomer at the Infrared Telescope Facility, which is operated by the University of Hawaii under contract NNH14CK55B with the National Aeronautics and Space Administration. This publication makes use of data products from the Two Micron All Sky Survey, which is a joint project oversity of Massachusetts and the Infrared Processing and Analysis Center/California Institute hnology, funded by the National Aeronautics and Space Administration and the National Science Foundation.   We made use of {\it Swift}/UVOT data reduced by P. J. Brown and released in the {\it Swift} Optical/Ultraviolet Supernova Archive (SOUSA). SOUSA is supported by NASA’s Astrophysics Data Analysis Program through grant NNX13AF35G.   

\vspace{5mm}
\facilities{ARIES:ST, ARIES:DFOT, CTIO:PROMPT, Las Cumbres Observatory (FLOYDS, Sinistro), Gemini:Gillett (GNIRS), IRTF (SpeX), Magellan:Baade (FIRE), Magellan:Clay (MIKE), ING:Newton (IDS), NTT (EFOSC2, SOFI), MMT (BCS), Keck:I (HIRES), Beijing:2.16m (BFOSC), NOT (ALFOSC), SOAR (Goodman), Asiago:Copernico (AFOSC), Asiago:Galileo (B\&C), VLT:Antu (FORS2), HCT (HFOSC), HET(LRS2), SO:Super-LOTIS, Bok (B\&C), Liverpool:2m (IO:O), REM}

\software{
astropy \citep{2013A&A...558A..33A,astropy},  {\sc lcogtsnpipe} \citep{Valenti16}, superbol \citep{supbol18}}

\bibliographystyle{aasjournal}
\bibliography{ms}

\appendix \label{appendix}

\section{Photometry}

\subsection{UV and Optical}

Photometric data for SN~2017gmr was obtained from a variety of telescopes (see Section~\ref{observations}), resulting in an extremely high cadence optical light curve (Figure~\ref{fig:fulllc}), as well as an early time {\it Swift} UV+optical light curve (Figure~\ref{fig:Ulc}).  We briefly describe the instrumentation and data reduction techniques here, although if a telescope+instrument combination is not specifically mentioned, the data was reduced in a `standard' way, including: image detrending (bias subtraction and flat fielding), cosmic ray removal, PSF or aperture photometry, along with flux calibration performed against standard catalogs (e.g. Landolt standard stars or the SDSS).  The full ground-based optical data set is presented in Table~\ref{tab:optphot}, while the {\it Swift} data is presented in Table~\ref{tab:swiftphot}.

%\starttable
\begin{deluxetable}{lcccc}[!h]
\tablecaption{SN~2017gmr Optical Photometry \label{tab:optphot}}
\tablehead{\colhead{MJD} & \colhead{Phase } & \colhead{Magnitude} & \colhead{Error} & \colhead{Telescope} \\
}
\startdata
\multicolumn{5}{c}{$U$}\\
\hline 
58000.276 & +1.19 & 14.44 & 0.05 & LCO-1m \\
58000.280 & +1.19 & 14.37 & 0.04 & LCO-1m \\
58000.358 & +1.27 & 14.11 & 0.02 & LCO-1m \\
58000.362 & +1.27 & 14.11 & 0.02 & LCO-1m \\
58000.631 & +1.54 & 13.90 & 0.03 & LCO-1m \\
58000.635 & +1.54 & 13.90 & 0.03 & LCO-1m \\
58000.984 & +1.89 & 13.87 & 0.05 & LCO-1m \\
58000.987 & +1.89 & 13.94 & 0.03 & LCO-1m \\
58001.115 & +2.03 & 13.89 & 0.03 & LCO-1m \\
58001.119 & +2.03 & 13.91 & 0.03 & LCO-1m \\
\hline 
\enddata
\tablecomments{Phases are reported with respect to an assumed explosion epoch of MJD 57999.09. Table 1 is published in its entirety in the machine-readable format. A portion is shown here for guidance regarding its form and content.}
\end{deluxetable}

\begin{deluxetable}{lccc}%[!t]
\tablecaption{SN~2017gmr {\it Swift} Photometry \label{tab:swiftphot}}
\tablehead{\colhead{MJD} & \colhead{Phase } & \colhead{Magnitude} & \colhead{Error} \\
}
\startdata
\multicolumn{4}{c}{$UVW2$}\\
\hline 
58001.626 & +2.5 & 13.80 & 0.07 \\
58002.767 & +3.7 & 14.05 & 0.07 \\
58003.565 & +4.5 & 14.26 & 0.20 \\
58008.442 & +9.3 & 15.26 & 0.08 \\
\hline 
\multicolumn{4}{c}{$UVM2$}\\
\hline 
58001.606 & +2.5 & 13.97 & 0.15 \\
58002.771 & +3.7 & 14.20 & 0.06 \\
58003.569 & +4.5 & 14.40 & 0.06 \\
58008.313 & +9.2 & 15.03 & 0.07 \\
\hline 
\multicolumn{4}{c}{$UVW1$}\\
\hline 
58001.622 & +2.5 & 13.60 & 0.07 \\
58002.764 & +3.7 & 13.71 & 0.08 \\
58003.562 & +4.5 & 13.78 & 0.07 \\
58008.441 & +9.3 & 14.27 & 0.20 \\
\hline 
\multicolumn{4}{c}{$u$}\\
\hline 
58001.623 & +2.5 & 13.58 & 0.05 \\
58002.766 & +3.7 & 13.60 & 0.05 \\
58003.563 & +4.5 & 13.62 & 0.05 \\
58008.442 & +9.3 & 13.70 & 0.05 \\
\hline 
\multicolumn{4}{c}{$b$}\\
\hline 
58001.624 & +2.5 & 14.81 & 0.30 \\
58002.766 & +3.7 & 14.70 & 0.14 \\
58003.563 & +4.5 & 14.65 & 0.12 \\
58008.442 & +9.3 & 14.62 & 0.10 \\
\hline 
\multicolumn{4}{c}{$v$}\\
\hline 
58001.596 & +2.5 & 14.52 & 0.08 \\
58002.769 & +3.7 & 14.43 & 0.07 \\
58003.567 & +4.5 & 14.51 & 0.07 \\
58008.312 & +9.2 & 14.25 & 0.17 \\
\enddata
\tablecomments{Phases are reported with respect to an assumed explosion epoch of MJD 57999.09.}
\end{deluxetable}

First, continued monitoring of SN~2017gmr was done by the DLT40 survey's two discovery telescopes, the PROMPT5 0.4-m telescope at Cerro Tololo Inter-American Observatory and the PROMPT-MO 0.4-m telescope at Meckering Observatory in Australia, operated by the Skynet telescope network \citep{Reichart05}.  The PROMPT5 telescope has no filter (`Open') while the PROMPT-MO telescope has a broadband `Clear' filter, both of which we calibrate to the Sloan Digital Sky Survey $r$ band \citep[see ][for further reduction details]{Tartaglia18}. 

Las Cumbres Observatory $UBVgri$-band data were obtained with the Sinistro cameras on the 1-m telescopes, through the Global Supernova Project. Using {\sc lcogtsnpipe} \citep{Valenti16}, a PyRAF-based photometric reduction pipeline, PSF fitting was performed. $UBV$-band data were calibrated to Vega magnitudes \citep{Stetson00} using standard fields observed on the same night by the same telescope. Finally, $gri$-band data were calibrated to AB magnitudes using the Sloan Digital Sky Survey \citep[SDSS,][]{sdssdr13}. Because the Las Cumbres data is the most comprehensive, and there are differences across the instrument/filter pairs, all other datasets were shifted by small amounts to match the Las Cumbres magnitudes in Figure \ref{fig:fulllc}. These values ranged between 0.05 and 0.15 magnitudes. The non-shifted values are all included in Table~\ref{tab:optphot}.

Optical photometry in the $BVRI$ bands was obtained at the 60/90cm Schmidt-telescopes at Konkoly Observatory; see \citet{Vinko12} for a description of the instrumentation and data reduction techniques.  Further, $BVRI$ photometry was obtained with the Super-LOTIS \citep[Livermore Optical Transient Imaging System;][]{slotis} 0.6 m telescope at Kitt Peak National Observatory; these data were reduced in a manner similar to that described in \citet{Kilpatrick16} and PSF photometry using standard IRAF procedures was then done on the resultant images. 

Data in the $BVugriz$ bands were taken with the IO:O imager on the Liverpool telescope, and were reduced using the standard IO:O pipeline; aperture photometry was performed using custom {\sc python} scripts and {\sc pyraf}. The data were shifted $+$0.17 mag in $B$ to match the Las Cumbres data. Data from the 1.30-m DFOT and 2.01 HCT telescopes were reduced as described in \citet{Dastidar19} performing PSF fitting photometry using {\sc DAOPHOT II} \citep{Stetson87}.  Instrumental magnitudes were converted to standard magnitudes using a set of local standard stars, and observations of either Landolt standard or SDSS fields.

The {\it Swift} UVOT analysis uses the pipeline of the {\it Swift} Optical/Ultraviolet Supernova Archive\footnote{http://swift.gsfc.nasa.gov/docs/swift/sne/swift\_sn.html} (SOUSA; \citealp{Brown_etal_2014}).  The method is based on that of \citet{Brown_etal_2009}, including subtraction of the host galaxy count rates, and uses the revised UV zeropoints and time-dependent sensitivity from \citet{Breeveld_etal_2011}.  For SN~2017gmr we do not have template images to subtract the underlying galaxy flux.  In this case, however, the largest contributor to the background is scattered/reflected light from the nearby bright star evident in Figure~\ref{fig:image}.  The reported UVOT magnitudes use a background position which to the eye approximated the brightness of the galaxy and halo at the SN position.  The errors have been conservatively increased to match the range of magnitudes measured with a variety of halo-free and bright halo regions.  The full {\it Swift} data set is presented in Table~\ref{tab:swiftphot} and is plotted in Figure~\ref{fig:Ulc}. 

\subsection{Near Infrared}

Raw NIR data from NOTCam was reduced using the NOTCam Quicklook reduction package and PSF photometry was then performed using standard IRAF procedures. The REM telescope is equipped with an optical and an IR camera, which observes simultaneously the same field, thanks to a dichroic placed before the telescope focal plane. IR images were corrected for dark current and flat fielded, and subsequently median stacked to obtain a background frame for each filter. The background-subtracted images were geometrically aligned and then stacked to obtain a final image for each filter, and the background in the locations of  SN~2017gmr was modeled with a low order polynomial surface and subtracted. The flux of the SN and the local sequence was measured through PSF fitting. For both instruments, photometric calibration was done using 2MASS stars in the field.  The resulting dataset can be found in Table~\ref{tab:NIRphot}.

\begin{table}
\begin{center}\begin{minipage}{3.3in}
      \caption{NIR Photometry}
\begin{tabular}{cccc}\hline\hline
  MJD    & $J$ & $H$ & $K$  \\
  \hline
\multicolumn{4}{c}{NOT}\\
  \hline
58005.5&	13.84 $\pm$ 0.05&	13.77 $\pm$ 0.07&	13.94 $\pm$ 0.12\\
58025.3&	13.13 $\pm$ 0.05&	12.91 $\pm$ 0.07&	12.70 $\pm$	0.12\\
58042.5&	13.08 $\pm$ 0.05&	12.84 $\pm$ 0.07&	12.86 $\pm$ 0.12\\
58121.1&	14.43 $\pm$	0.05&	14.18 $\pm$ 0.07&	14.19 $\pm$ 0.12\\
58140.5&	14.72 $\pm$	0.05&	14.63 $\pm$ 0.07&	14.67 $\pm$ 0.12\\
58165.8&	15.27 $\pm$	0.05&	15.19 $\pm$ 0.07&	15.23 $\pm$ 0.12\\
\hline
\multicolumn{4}{c}{REM}\\
\hline
58012.26&	13.23 $\pm$ 0.03&	13.09 $\pm$ 0.03&	12.88 $\pm$ 0.06\\
58017.32&	13.23 $\pm$ 0.03&	13.05 $\pm$ 0.03&	12.87 $\pm$	0.06\\
58021.31&	13.26 $\pm$ 0.03&	12.96 $\pm$ 0.03&	12.84 $\pm$ 0.03\\
58022.31&	14.21 $\pm$	0.03&	12.97 $\pm$ 0.03&	12.77 $\pm$ 0.22\\
58027.25&	13.03 $\pm$	0.04&	12.81 $\pm$ 0.04&	--\\
58073.12&	13.09 $\pm$	0.04&	12.79 $\pm$ 0.05&	--\\
58083.06&	13.23 $\pm$	0.03&	12.98 $\pm$ 0.04&	--\\
58093.07&	13.59 $\pm$	0.03&	13.35 $\pm$ 0.04&	--\\
58109.21 &	14.28 $\pm$	0.04&	14.05 $\pm$ 0.05&	--\\
\hline
\end{tabular}\label{tab:NIRphot}
\end{minipage}\end{center}
\end{table}

\section{Spectroscopy}
 \subsection{Optical Spectroscopy}

A high cadence spectral sequence of SN~2017gmr was taken with low, medium and high -resolution instrumentation throughout the rise, plateau,  and fall from plateau of the supernova.  A log of these observations can be found in Table~\ref{tab:optspec}.  These spectra were reduced using standard techniques, including bias subtraction, flat fielding, cosmic ray rejection, local sky subtraction and extraction of one-dimensional spectra. Most observations had the slit aligned along the parallactic angle to minimize differential light losses.  Flux calibration was done with standard star observations, and most spectra were rescaled to match the photometric light curve at a given epoch.  We discuss some details of the spectroscopic reductions below, but if a particular telescope+instrument combination is not mentioned, it was reduced in a standard way as described above.

Las Cumbres optical spectra were taken with the FLOYDS spectrographs mounted on the 2m Faulkes Telescope North and South at Haleakala, USA and Siding Spring, Australia, respectively, through the Global Supernova Project. A 2$\arcsec$ slit was placed on the target at the parallactic angle. One-dimensional spectra were extracted, reduced, and calibrated following standard procedures using the FLOYDS pipeline \citep{Valenti14}.HIRES spectra were reduced using the MAuna Kea Echelle Extraction (MAKEE) data reduction package\footnote{http://spider.ipac.caltech.edu/staff/tab/makee/} (written by T. Barlow). MIKE spectra were
reduced using the latest version of the MIKE pipeline\footnote{http://code.obs.carnegiescience.edu/mike/} (written by D. Kelson).

%\subsection{Infrared}
\subsection{NIR Spectroscopy}

A sequence of NIR spectra of SN~2017gmr were also taken, and are logged in Table~\ref{tab:nirspec}.  All NIR spectra were taken using a classical ABBA technique, dithering the object along the slit in order to facilitate good sky subtraction.  Further, the slit was oriented along the parallactic angle to minimize slit losses due to atmospheric differential refraction \citep{filippenko82}.  In all cases, an A0V star was observed either before or after the science observations in order to correct for telluric absorption and flux calibrate the data, following the prescriptions of \citet{Vacca03}.

%Three spectra were taken with the Gemini Near-Infrared Spectrograph (GNIRS) at Gemini North Observatory \citep{Elias06} 
Gemini/GNIRS data was taken in cross-dispersed mode with the 0.675$\arcsec$ slit, yielding continuous wavelength coverage from 0.8 to 2.5 $\mu$m and an $R\sim$1000.  These data were reduced with the {\sc XDGNIRS} pipeline provided by Gemini Observatory, as described in \citet{Sand16,Hsiao19}.

%Three additional NIR spectra were taken with SpeX \citep{Rayner03} on the NASA Infrared Telescope Facility (IRTF) 
IRTF spectra were taken in SXD mode and the 0.5 arcsec slit, yielding wavelength coverage from $\sim$0.8--2.4 $\mu$m and $R\sim$1200.  These data were reduced using the publicly available Spextool software package \citep{Cushing04}, as described in \citet{Hsiao19}.

%Need a section on NTT/SOFI spectra
Two NIR spectra were taken with the Son OF ISAAC (SOFI) spectrograph mounted on the NTT telescope \citep{SOFI}, using both the Blue and Red grisms, giving a broad wavelength coverage of 0.9--2.4 $\mu$m.  The SOFI spectra were taken as part of the ePESSTO program, and were reduced as described in \citet{Smartt15}.

Finally, a single FIRE spectrum was taken using the high throughput prism mode with a 0.6 arcsec slit, giving continuous wavelength coverage from 0.8--2.5 $\mu$m and a resolution of $R$$\sim$500 in the $J$ band.  The spectrum was reduced with the purpose-built {\sc firehose} pipeline \citep{Simcoe13} as described in detail in \citet{Hsiao19}.

 \startlongtable
 \begin{deluxetable*}{lcccccc}
\tablecaption{Optical Spectroscopy of SN~2017gmr \label{tab:optspec}}
\tablehead{ \colhead{UT Date}    &\colhead{MJD}& \colhead{Phase}    &\colhead{Telescope+}   & \colhead{R}&  \colhead{Exposure Time}  \\ 
   \colhead{(y-m-d)}    &\colhead{} & \colhead{(days)} & \colhead{Instrument}  &\colhead{$\lambda$/$\Delta_{\lambda}$}   & \colhead{(s)}   \\  }
\startdata
2017-09-04 & 58000.57 & 1.5 &  FTS+FLOYDS & 500 & 2700 \\
2017-09-04 & 58000.59 & 1.5 &  Keck+HIRES & 50000 & 3$\times$900 \\
2017-09-05 & 58001.34 & 2.2 &  SOAR+Goodman &500 & 900 \\
2017-09-05 & 58001.43 & 2.3 & HET+LRS2B & 1100 & 1000 \\
2017-09-05 & 58001.62 & 2.5 &  FTS+FLOYDS & 500& 2700 \\
2017-09-06 & 58002.19 & 3.1 &  INT+IDS& 300& 2$\times$900 \\
2017-09-08 & 58004.19 & 5.1 &  INT+IDS& 300& 1200 \\
2017-09-08 & 58004.35 & 5.3 &  Mag+IMACS&4000 & 300 \\
2017-09-08 & 58004.44 & 5.4 & HET+LRS2B & 1100 & 1000\\
2017-09-08 & 58004.55 & 5.5 &  HCT+HFOSC&350 & 2$\times$1200 \\
2017-09-08 & 58004.56 & 5.5 &  FTS+FLOYDS& 500  & 2700 \\
2017-09-10 & 58006.19 & 7.1 &  NOT+ALFOSC &300 & 900 \\
2017-09-10 & 58006.76 & 7.7 &  FTS+FLOYDS& 500 & 2700 \\
2017-09-11 & 58007.31 & 8.2 &  NTT+EFOSC2&200 & 600 \\
2017-09-11 & 58007.41 & 8.3 & HET+LRS2B & 1100 & 1000\\
2017-09-12 & 58008.50 & 9.4 &  FTN+FLOYDS& 500 & 2700 \\
2017-09-12 & 58008.50 & 9.4 &  Bok+BC & 700 & 3$\times$120 \\
2017-09-15 & 58011.47 & 12.4 &  FTN+FLOYDS& 500 & 2700 \\
2017-09-16 & 58012.13 & 13.0 &  NOT+ALFOSC&300 & 2$\times$300 \\
2017-09-16 & 58012.69 & 13.6 &  FTS+FLOYDS& 500 & 2700 \\
2017-09-18 & 58014.53 & 15.4 &  FTN+FLOYDS& 500 & 2700 \\
2017-09-19 & 58015.78 & 16.7 &  BAO+BFOSC &500 & 2400 \\
2017-09-22 & 58018.06 & 19.0 &  NOT+ALFOSC&300 & 2$\times$300 \\
2017-09-22 & 58018.69 & 19.6 &  BAO+BFOSC&500 & 2400 \\
2017-09-25 & 58021.42 & 22.3 &  FTN+FLOYDS& 500 & 2700 \\
2017-09-26 & 58022.78 & 23.7 &  BAO+BFOSC&500 & 2400 \\
2017-09-27 & 58024.00 & 24.9 &  NOT+ALFOSC&300 & 2$\times$300 \\
2017-09-29 & 58025.40 & 26.4 &  Bok+BC & 700 & 3$\times$600 \\
2017-10-03 & 58029.53 & 30.4 &  FTN+FLOYDS& 500 & 2700 \\
2017-10-05 & 58031.06 & 32.0 &  NOT+ALFOSC&300 & 600 \\
2017-10-11 & 58037.13 & 38.0 &  NOT+ALFOSC&300 & 600 \\
2017-10-11 & 58037.40 & 38.3 &  Bok+BC & 700 & 3$\times$240 \\
2017-10-17 & 58043.01 & 43.9 &  Asiago182+AFOSC &300& 2$\times$1200 \\
2017-10-17 & 58043.25 & 44.2 &  VLT+FORS2&500 & 274+343 \\
2017-10-18 & 58044.74 & 45.7 &  HCT+HFOSC&350 & 2$\times$1200 \\
2017-10-18 & 58044.99 & 45.9 &  Asiago122+BC &700 & 3$\times$1800 \\
2017-10-21 & 58047.26 & 48.2 &  NTT+EFOSC2&200 & 900 \\
2017-10-22 & 58048.54 & 49.5 &  FTS+FLOYDS&500 & 2700 \\
2017-10-24 & 58050.10 & 51.0 &  VLT+FORS2 &500 & 2$\times$299 \\
2017-10-27 & 58053.52 & 54.4 &  FTS+FLOYDS& 500 & 2700 \\
2017-10-27 & 58053.70 & 54.6 &  MMT+BCH& 3900 & 3$\times$300 \\
2017-10-28 & 58054.37 & 55.3 &  Bok+BC & 700 & 3$\times$240 \\
2017-10-29 & 58055.28 & 56.3 & HET+LRS2B & 1100 & 1000\\
2017-11-01 & 58058.45 & 59.4 &  FTS+FLOYDS& 500 & 2700 \\
2017-11-02 & 58059.80 & 60.7 &  HCT+HFOSC&350 & 2$\times$1800 \\
2017-11-03 & 58060.66 & 61.6 &  HCT+HFOSC &350& 2$\times$1800 \\
2017-11-05 & 58062.79 & 63.7 &  HCT+HFOSC&350 & 2$\times$1200 \\
2017-11-06 & 58063.48 & 64.4 &  FTN+FLOYDS& 500 & 2700 \\
2017-11-06 & 58063.79 & 64.7 &  HCT+HFOSC&350 & 2$\times$1200 \\
2017-11-10 & 58067.09 & 68.0 &  NOT+ALFOSC&300 & 2$\times$300 \\
2017-11-10 & 58067.24 & 68.2 &  VLT+FORS2&500 & 329 \\
2017-11-10 & 58067.79 & 68.7 &  HCT+HFOSC&350 & 2$\times$1200 \\
2017-11-13 & 58070.63 & 71.5 &  BAO+BFOSC&500 & 3000 \\
2017-11-19 & 58076.39 & 77.3 &  FTN+FLOYDS& 500 & 1800 \\
2017-11-19 & 58076.73 & 77.6 &  HCT+HFOSC &350 & 2$\times$1800 \\
2017-11-20 & 58077.31 & 78.2 &  MMT+BCH& 3900 & 3$\times$300 \\
2017-11-23 & 58080.72 & 81.6 &  HCT+HFOSC&350 & 2$\times$1200 \\
2017-11-26 & 58083.95 & 84.9 &  Asiago122+BC&700 & 4$\times$1800 \\
2017-11-29 & 58086.57 & 87.5 &  HCT+HFOSC&350 & 2$\times$1200 \\
2017-12-02 & 58089.60 & 90.5 &  HCT+HFOSC&350 & 2$\times$1800 \\
2017-12-03 & 58090.91 & 91.8 &  Asiago122+BC&700 & 4$\times$1800 \\
2017-12-04 & 58091.38 & 92.3 &  FTN+FLOYDS& 500 & 1800 \\
2017-12-07 & 58094.04 & 94.9 &  NOT+ALFOSC&300 & 600 \\
2017-12-09 & 58096.83 & 97.7 &  Asiago122+BC&700 & 4$\times$1800 \\
2017-12-10 & 58097.26 & 98.2 &  Bok+BC & 700 & 3$\times$1200 \\
2017-12-12 & 58099.15 & 100.1 &  VLT+FORS2&500 & 2$\times$329 \\
2017-12-13 & 58100.12 & 101.0 &  VLT+FORS2&500 & 329 \\
2017-12-14 & 58101.04 & 102.0 &  VLT+FORS2&500 & 329 \\
2017-12-15 & 58102.46 & 103.4 &  FTS+FLOYDS& 500 & 3600 \\
%2017-12-18 & 58105.72 & 106.6 &  HCT+HFOSC&350 & 2$\times$1500 \\
2017-12-18 & 58105.8 & 106.7 &  Asiago122+BC&700 & 4$\times$1800 \\
2017-12-20 & 58107.8 & 108.7 &  Asiago122+BC&700 & 4$\times$1800 \\
2017-12-21 & 58108.11 & 109.0 &  VLT+FORS2 &500& 3$\times$329 \\
2017-12-21 & 58108.55 & 109.5 &  BAO+BFOSC&500 & 3000 \\
2017-12-22 & 58109.11 & 110.0 &  VLT+FORS2&500 & 2$\times$329 \\
2017-12-23 & 58109.8 & 110.8 &  Asiago122+BC&700 & 4$\times$1800 \\
2017-12-23 & 58110.04 & 111.0 &  VLT+FORS2&500 & 329 \\
2017-12-23 & 58110.23 & 111.1 &  MMT+BCH & 3900 &3$\times$300 \\
2017-12-29 & 58116.59 & 117.5 &  HCT+HFOSC&350 & 2$\times$1500 \\
2018-01-05 & 58123.45 & 124.4 &  FTS+FLOYDS& 500 & 3600 \\
2018-01-07 & 58126.42 & 127.3 &  BAO+BFOSC&500 & 3000 \\
2018-01-09 & 58128.82 & 129.7 &  Asiago182+AFOSC &300& 2$\times$1200 \\
2018-01-14 & 58132.05 & 133.0 &  VLT+FORS2 &500& 329 \\
2018-01-14 & 58132.56 & 133.5 &  HCT+HFOSC&350& 2$\times$1500 \\
2018-01-15 & 58133.07 & 134.0 &  VLT+FORS2&500 & 329 \\
2018-01-16 & 58134.12 & 135.0 &  VLT+FORS2 &500& 329 \\
2018-01-16 & 58134.26 & 135.2 &  FTN+FLOYDS& 500 & 3600 \\
2018-01-17 & 58135.05 & 136.0 &  VLT+FORS2&500 & 329 \\
2018-01-18 & 58136.05 & 137.0 &  VLT+FORS2&500 & 329 \\
2018-01-17 & 58136.46 & 137.4 &  BAO+BFOSC&500 & 3300 \\
2018-01-19 & 58137.07 & 138.0 &  VLT+FORS2&500 & 2$\times$329 \\
2018-01-20 & 58137.89 & 138.8 &  NOT+ALFOSC&300 & 600 \\
2018-01-22 & 58141.47 & 142.4 &  BAO+BFOSC&500 & 3000 \\
2018-01-24 & 58142.15 & 143.1 &  Bok+BC& 700 & 3$\times$600 \\
2018-02-05 & 58154.09 & 155.0 &  MMT+BCH & 3900& 3$\times$900 \\
2018-02-12 & 58161.22 & 162.1 &  FTN+FLOYDS& 500 & 3600 \\
2018-01-20 & 58163.86 & 164.7 &  NOT+ALFOSC&300 & 600 \\
2018-07-12 & 58311.37 & 312.3 &  Magellan+MIKE &40000& 3$\times$1200 \\
\hline
\enddata
 \tablecomments{Phases are reported with respect to an explosion epoch of 57999.09}
 \end{deluxetable*}

 \begin{table*}
 \begin{center}%\begin{minipage}{3.3in}
      \caption{NIR Spectroscopy of SN~2017gmr}
 \begin{tabular}{@{}lcccccc}\hline\hline
  UT Date    &MJD&Phase &Telescope    & $\lambda$ range  & Exp Time   \\ 
      
   (y-m-d)    & &&Instrument&     $\mu$m    & (s)   \\   \hline
% %%
2017-09-06 &58002.46 &+3.4     & Gemini/GNIRS &      0.82--2.4 & 14$\times$120   \\
2017-09-10 & 58005.57 &+6.5 & IRTF/SpeX &  0.82--2.4 & 8$\times$150 \\
2017-09-16 & 58012.45 &+13.4 &  Gemini/GNIRS & 0.82--2.4 & 20$\times$150\\
2017-10-20 & 58046.17 & +47.1 &NTT/SOFI &  0.9--2.4 & 12$\times$125 \\
2017-10-28 & 58054.31 &+54.2 & IRTF/SpeX &  0.82--2.4 & 12$\times$150 \\
2017-11-28 & 58085.14 &+86.1 & NTT/SOFI &  0.9--2.4 & 8$\times$120 \\
2017-12-11 & 58098.30 &+99.2 & IRTF/SpeX &  0.82--2.4 & 20$\times$150 \\
2018-01-20 & 58138.23 &+139.1 & Gemini/GNIRS &  0.8--2.4 & 20$\times$120 \\
2018-01-31 & 58149.02 &+149.9 & Magellan/FIRE &  0.8--2.4 & 12$\times$127\\
 \hline
 \end{tabular}\label{tab:nirspec}
 \tablenotetext{}{NOTES:Phase calculated with respect to our assumed explosion epoch, MJD = 57999.09.}
%\figcaption{Phase calculated with respect to our assumed explosion epoch, JD=2458000.266.}
 %MJD_expl =  57999.766zXZ
% \end{minipage}
 \end{center}
 \end{table*}

\end{document}